\documentclass[aps,prb,reprint,groupedaddress, showpacs]{revtex4-2}
\usepackage{graphicx,graphics}
\usepackage[table]{xcolor}
\usepackage{epstopdf}
\usepackage{amsmath,amssymb,amsfonts,stmaryrd}
\usepackage{latexsym,verbatim}
\usepackage{bm}
\usepackage{bbold}
\usepackage{xcolor}
\usepackage[normalem]{ulem}
\usepackage[breaklinks=true,colorlinks,citecolor=blue,linkcolor=blue,urlcolor=blue]{hyperref}
\usepackage[capitalise]{cleveref}
\usepackage[abs]{overpic}
\usepackage{textcomp} 
\usepackage{mathtools}
\usepackage{braket}
\usepackage{comment}
\usepackage{tcolorbox}
\usepackage{soul}

\begin{document}
 
\title{Vibrational response functions for multidimensional electronic spectroscopy: \\ 
from Duschinsky rotations to multimode squeezed coherent states}

\author{Frank Ernesto Quintela Rodriguez}
\affiliation{Università di Modena e Reggio Emilia, I-41125 Modena, Italy}
\author{Filippo Troiani$^*$}
\affiliation{Centro S3, CNR-Istituto di Nanoscienze, I-41125 Modena, Italy}
\affiliation{\rm $^*$Author to whom correspondence should be addressed: filippo.troiani@nano.cnr.it}


\begin{abstract}
Multidimensional spectroscopy unveils the interplay of nuclear and electronic dynamics, which characterizes the ultrafast dynamics of various molecular and solid-state systems. 
In a class of models widely used for the simulation of such dynamics, field-induced transitions between electronic states result in linear transformations (Duschinsky rotations) between the normal coordinates of the vibrational modes. 
Here we present an approach for the calculation of the response functions, based on the explicit derivation of the vibrational state. This can be shown to coincide with a multimode squeezed coherent state, whose expression we derive within a quantum-optical formalism,
and specifically by the sequential application to the initial state of rotation, displacement and squeeze operators. The proposed approach potentially simplifies the numerical derivation of the response functions, avoiding the time integration of the Schr\"odinger equation, the Hamiltonian diagonalization, and the sum over infinite vibronic pathways. Besides, it quantitatively substantiates in the considered models the intuitive interpretation of the response functions in terms of the vibrational wave packet dynamics.
\end{abstract}

\date{\today}

\maketitle

\section{Introduction}

Coherent multidimensional spectroscopy employs ultra-short laser pulses to probe the photo-physical dynamics in the femtosecond regime \cite{Hamm11a,Maiuri2020}.
The dependence of the observed polarization on the time delays between the pulses allows one to correlate the system's resonant frequencies and to discriminate distinctive features in the quantum dynamics \cite{Mukamel95a,Biswas22a}.
This technique has been widely applied to the study of diverse ultrafast physical processes, including, among others \cite{kennis2007ultrafast,Christensson2012,lloyd20212021}, charge separation \cite{croce2018light}, internal conversion \cite{moretti2020ultrafast}, generation of polaron-pairs \cite{DeSio16a}, and electron transfer \cite{Cao2020,Rafiq2021}.
However, the interpretation of the multidimensional spectra is not immediate, and requires the aid of numerical simulations. These can complement the observable quantities with information on the potential energy surfaces (PESs) \cite{anda2018two,Weakly2021} and, where needed, on the non-adiabatic couplings between electronic and nuclear degrees of freedom \cite{zhang2016effects,Bizimana2017a,Kundu22a,Tiwari}.

In order to simulate the multidimensional spectra, the systems can typically be modelled by a few electronic states, coupled to a set of independent harmonic oscillators. These are derived from the normal mode analysis of the PESs, performed within electronic structure calculations \cite{kuhn2019frenkel}. 
A full ab initio simulation of the spectra becomes in fact impractical as the size of the system increases and contributions from highly excited-states show up in the measured signal \cite{prasad2021computational,young2004computational,begusic2021finite}.
Model Hamiltonians that are quadratic in the momentum and coordinate operators, usually referred to as generalized Brownian oscillator models  \cite{Mukamel95a,Fidler13a,Zuehlsdorff2020},
account for the dominant (lower-order) contributions to the interaction potentials and have been extensively used in the simulation of experiments \cite{Mancal10a,seiler2019two,shen2022simulation}.
In adiabatic models, the vibrational and electronic degrees of freedom factorize at the level of eigenstates \cite{breuer2002theory,Butkus12a,Cina2016,Le21a}, but can mix in the dynamics triggered by the laser pulses.
In particular, in the Franck-Condon approximation, the laser pulses induce transitions between electronic states and simultaneously launch vibrational wave packets in the excited-state PES. Here they freely propagate during the time delays, giving rise to a modulation of the multidimensional electronic spectra. Specific features of such modulation can be typically traced back to the displacement of the equilibrium position of the PESs\cite{Mukamel95a}, to their electronic-state dependent curvature \cite{engel2013}, and to the Duschinsky multi-mode mixing \cite{Doktorov77a}.

These Hamiltonian parameters contribute to the physical and spectral features in many different ways. 
It is known that a temperature dependence of the absorption line shape {\color{black} results from a difference between the ground and excited state vibrational frequencies}  \cite{engel2013}. 
In 2D spectroscopy, the electronic-state dependent frequencies can be used to unambiguously separate signals from excited and ground-state wave-packets \cite{Caram2012}.
Large differences in the curvatures of the PESs are also a notable feature of molecular photoswitches, which are currently studied by a wide variety of potential applications in nanotechnology \cite{smith2021modelling,pianowski2019recent}.
The Duschinsky mixing of normal modes can affect the transition rates for electron transfer \cite{sando2001large,tang2003effects}, internal conversion \cite{peng2007excited} and fluorescence rates \cite{de2018theoretical,niu2010theory},
as well as producing the cooling of the vibrational population in the excited state \cite{ianconescu2004photoinduced}.
Their spectral signatures have been be resolved by transient–absorption measurements \cite{arpin2022signatures}, electronic-vibrational spectroscopy \cite{courtney2015two,Fumero2020}, and 2D electronic spectroscopy \cite{Zuehlsdorff2020}.

Despite the simplifications introduced in the harmonic model, the calculation of the response functions might {\color{black} turn out to be} complicated, due to the inherent multilevel and multi-mode mixing present in the model. Exact analytical results for the linear response have been obtained from the quadratic propagator derived in the Feynman’s path integral formulation \cite{Baiardi2013a}. 
Arbitrary orders of the nonlinear response can in principle be calculated from the known eigenstates of the harmonic model, but even there technical complications arise from the calculation of the Franck-Condon factors and from the need to include, in principle, an infinite number of vibronic pathways \cite{chang2008new,sattasathuchana2020generalized}. Analytical calculations of the nonlinear response follow the cumulant expansion of the energy gap operator between electronic states.
The exact third-order response is known for the two-level system with uniform PES curvatures and no Duschinsky mixing \cite{Mukamel95a}, and has been extended to multi-levels systems coupling an arbitrary system Hamiltonian to a harmonic bath \cite{sung2001four,sung2003optical}.
This is possible because quantum correlation functions beyond the second order vanish when the energy gap follows a Gaussian statistics, so that the truncation of the expansion at second order is exact.
However, a non-uniform PES curvature and the Duschinsky mixing introduce non-Gaussian fluctuations, requiring that higher-order cumulants be taken into account \cite{Zuehlsdorff2020,zuehlsdorff2019optical,zuehlsdorff2021vibronic}.
The third-order response functions for harmonic oscillators with non-uniform PESs curvatures have been calculated in the third-order cumulant expansion \cite{engel2013}, and later extended by using the link between classical and quantum correlation functions \cite{Zuehlsdorff2020}. 
Alternative numerical strategies are based on the integration of the master equation, for example within the Lindblad or Redfield theories \cite{Rozzi18a,shah2022qudpy,rose2021efficient,rose2021automatic}, on the hierarchical equations of motion \cite{gelin2022equation} or an approximate dynamics \cite{jain2019simple}.

Here we follow a different route in the calculation of the vibrational response functions, which allows the calculation of arbitrary-order contributions in systems with arbitrary numbers of electronic levels. The approach is based on the explicit derivation of the vibrational state corresponding to each electronic pathways, {\it i.e.} to each sequence of electronic states, resulting from the laser-induced transitions. Besides providing the explicit expression of the vibrational states and a clear physical picture underlying the vibrational-induced modulation of the response functions, this approach does not require the counting over all the vibronic eigenstates of the full Hamiltonian, and thus circumvents the calculation of the Franck-Condon factors. The presented derivation employs a quantum optics formalism, and specifically the decomposition of the time-evolution generated by the vibrational Hamiltonian in terms of the rotation, displacement and squeeze operators \cite{Xin89a,Xin90a,Doktorov75a,Huh15a}. From this decomposition it follows that the squeezed-coherent character of the vibrational state is preserved throughout its time evolution. This is particularly relevant, because the vibrational ground state is a particular case of a squeezed coherent state, and coherent states form an over-complete basis, in terms of which any initial state can be decomposed. 

These results generalize those obtained along the same lines for the the linearly displaced harmonic oscillator model, where the electronic-state dependence of the vibrational frequencies and mode mixing are not present \cite{Quintela2022a}. While in that case closed analytical expressions could be derived for the vibrational response functions, here we provide iterative solutions for the quantities that define the multimode squeezed coherent states. The approach is applied for illustrative purposes to systems with two electronic levels and up to two vibrational modes. These examples clearly and intuitively show the connection between the dynamics of the vibrational wave packet(s) and the time dependence of the vibrational response functions.

The remainder of the paper is organized as follows. In Sect. II we introduce the approach by discussing the case of a single vibrational mode, with electronic-state dependent frequency. Section III presents the generalization of these results to the multimode case, where the mixing between the modes results from Duschinsky rotations. In Sect. IV we draw our conclusions. Background information is reported in Appendix \ref{app:a} and \ref{app:b} for the single- and multi-mode cases, respectively.

\section{Single vibrational mode with electronic-state dependent frequency}

In order to illustrate the procedure more smoothly, we start by discussing the simpler case of a single vibrational mode. The more general multimode case requires mathematically nontrivial, but conceptually straightforward, generalizations of the steps discussed in the present Section, and the inclusion of the Duschinsky rotations. These generalizations are discussed in the following Section.

\subsection{Definition of the model and of the approach}

The considered model consists of $N_e$ electronic levels and a vibrational mode. This is specifically given by a harmonic oscillator, whose origin and frequency depend on the electronic state. The corresponding Hamiltonian reads:
\begin{align}\label{eq:ham1}
  H = \sum_{\lambda=0}^{N_e} |\lambda\rangle\langle\lambda| \otimes \left( \epsilon_\lambda + H_{v,\lambda} \right),
\end{align}
where the vibrational components $H_{v,\lambda}$ are given by:
\begin{gather}\label{eq:hamvib}
    H_{v,\lambda} 
    \!=\! \omega_\lambda [(a_\lambda^\dagger\!+\!\delta_\lambda)(a_\lambda\!+\!\delta_\lambda)\!+\!1/2] .
\end{gather}
The displacements $\delta_\lambda$ are assumed to be real. Without loss of generality, we set for the ground state $\delta_0=0$. Given the dependence of the vibrational frequency on the electronic state, and thus on the pathway that specifies the contribution to the response function, the constant term $1/2$ in $H_{v,\lambda}$ cannot be neglected. As to the Hamiltonian eigenstates, these can always be written in a factorized form, given by the product of the electronic state $|\lambda\rangle$, and of the Fock state $|n_\lambda\rangle$, where $H_{v,\lambda}|n_\lambda\rangle = (\hbar\omega_\lambda+1/2) |n_\lambda\rangle$.

\begin{figure}[h]
\centering
\includegraphics[width=0.35\textwidth]{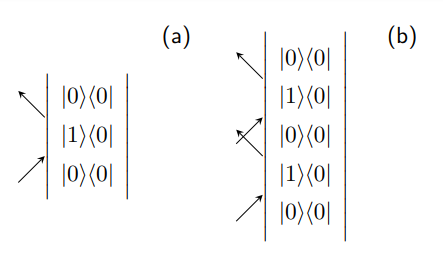}
\caption{Double-sided Feynman diagrams corresponding to the (a) first- and (b) third-order response functions of a system with two electronic levels, {\color{black} where the interactions with the field all occur on the left side}. The {\color{black} third-order diagram can be related to} the non-rephasing part of the ground-state bleaching contribution {\color{black} ($\lambda_1=1$, $\lambda_2=0$, and $\lambda_3=1$)}. As required in the present approach, the diagrams only specify the electronic pathway $p_e$.} 
\label{fig:0}
\end{figure}

{\color{black}
The response functions that are used in the simulation of multidimensional spectroscopy correspond to multi-time correlations function. In the following, we specifically refer to first- and third-order response functions, which are given, respectively, by \cite{Mukamel95a,Hamm11a}:
\begin{gather}\label{eq:sdd1}
    R^{(1)} =(i/\hbar) \langle \mu (t_1),[\mu,\rho_0]\rangle ,
\end{gather}
and
\begin{gather}\label{eq:sdd}
    R^{(3)} =(i/\hbar)^3 \langle \mu (t_{123}),[\mu (t_{12}),[\mu (t_1),[\mu,\rho_0]]]\rangle ,
\end{gather}
where $t_{123}\equiv t_1+t_2+t_3$ and $t_{ij}\equiv t_i+t_j$. The density operator $\rho_0$ defines the initial state of the system, which is hereafter assumed to coincide with its ground state, while 
$\mu(t)=U^\dagger (t)\,\mu\, U(t)$ is the dipole operator in the Heisenberg representation, with $U(t)=e^{-iHt/\hbar}$ and $\mu\equiv\mu (0)$. Within the Franck-Condon approximation, the system undergoes sudden transitions between electronic states, induced by the electric field, that leave the vibrational state unchanged. Correspondingly, the dipole operator coincides with the identity in the vibrational space.}

The response functions can be written as the sum over a number of contributions, one for each pathway \cite{Mukamel95a,Hamm11a}. In particular, we refer hereafter to the pathways (double sided Feynman diagrams) where the interactions with the field all occur on the left: all other cases can be derived from this one by suitably redefining the waiting times {\color{black} (see Appendix \ref{app:z} and Refs. \onlinecite{PhysRevE.47.118,Troiani23a})}. For an Hamiltonian such as the one defined by Eqs. (\ref{eq:ham1}-\ref{eq:hamvib}), there are in principle infinite vibronic pathways $p$, each one corresponding to a different sequence of transitions between electron-vibrational eigenstates $|\lambda,n_\lambda\rangle$: 
\begin{gather}\label{eq:vp}
p:\,|\lambda_1,n_{\lambda_1}\rangle \longrightarrow |\lambda_2,n_{\lambda_2}\rangle\dots\longrightarrow |\lambda_M,n_{\lambda_M}\rangle .
\end{gather}

Here we follow an alternative approach, allowed by the adiabatic character of the coupling between electronic and vibrational degrees of freedom \cite{Quintela2022a}. This consists in factorizing {\color{black} --- for each electronic pathway $p_e$ ---} the response function into the product of an electronic and a vibrational component,
\begin{gather}
    R_{\Lambda}^{(M)}({\bf t}) = R_{\Lambda}^{(e,M)}({\bf t})\,R_{\Lambda}^{(v,M)}({\bf t})\ \ \ [{\bf t}=( t_1,\dots,t_M)],
\end{gather}
and deriving the time evolution of the latter component, being $p_e$ {\color{black} [or, equivalently, $\Lambda=(\lambda_1,\dots,\lambda_M)$]} defined exclusively by the sequence of electronic states
\begin{gather}\label{eq:ep}
p_e:\,|\lambda_1\rangle \longrightarrow |\lambda_2\rangle\dots\longrightarrow |\lambda_M\rangle .
\end{gather}
Not including any dependence on the vibrational quantum numbers $n_{\lambda_k}$, the electronic pathways ($p_e$) are infinitely less numerous than the vibronic ones ($p$). The role played by the vibrational degrees of freedom is entirely captured by the electronic-pathway dependent evolution of the vibrational state. This is generated by an Hamiltonian that is piecewise constant:
\begin{gather}\label{eq:51}
    |\psi_{ket,\Lambda}\rangle = e^{-i H_{v,\lambda_M}t_M/\hbar} \dots e^{-i H_{v,\lambda_1}t_1/\hbar} |0_0\rangle,
\end{gather}
being {\color{black} $|0_0\rangle$ the ground vibrational state of the ground state harmonic oscillator Hamiltonian $H_{v,0}$}. 
The vibrational component of the response function is given by the overlap between these two states, \begin{gather}\label{eq:52}
    R_{\Lambda}^{(v,M)} ({\bf t}) = \langle \psi_{bra} | \psi_{ket,\Lambda} \rangle 
\end{gather}
where the former one is given by the ground vibrational state, $|\psi_{bra}\rangle = e^{-i\omega_0\sum_{k=1}^M t_k/2}|0_0\rangle$.
The {\color{black} $p_e$-dependent} vibrational state $| \psi_{ket,\Lambda} \rangle$ is given, for the present Hamiltonian, by a squeezed coherent state, and can thus be specified by one real and two complex numbers. 

\subsection{Rotation, displacement, and squeeze operators\label{subsec:1}}

If the harmonic oscillator is initialized in a squeezed coherent state, it remains in a squeezed coherent state at all times. This results from the fact that each time-evolution operator $e^{-i H_{v,\lambda}t/\hbar}$ can be written as the product of a squeeze, a displacement, and a rotation operator \cite{Doktorov77a,Xin89a}, and that the application of each of these operators to a squeezed coherent state gives rise to another squeezed coherent state.

\subsubsection{Definition of the operators\label{subsecY}}

For the reader's convenience, we report hereafter the definitions of these operators \cite{Gerry04a,Scully97a}. The rotation operator can be expressed as an exponential function of the number operator $a^\dagger a$:
\begin{gather}
    R(\phi) = \exp\left(i\phi\, a^\dagger a\right) ,
\end{gather}
where $\phi$ is real and defines the rotation angle. The time evolution operator generated by the undisplaced oscillator Hamiltonian $H_{v,0}$ corresponds to a rotation operator with $\phi=-\omega_{0}t$, multiplied by a phase factor $e^{-i\omega_0 t/2}$. 

Also the displacement operator can be written as an exponential function of the creation and annihilation operators, and specifically as:
\begin{gather}\label{A03}
    D(\beta) = \exp\left(\beta a^\dagger - \beta^* a\right) ,
\end{gather}
where $\beta$ is a complex number. In the absence of a frequency change, {\it i.e.} if $\omega_\lambda=\omega_0$ for any $\lambda$, the time evolution operator $e^{-iH_{v,\lambda}t/\hbar}$ can be written as a combination of a rotation and two displacement operators \cite{Quintela2022a}, times a phase factor $e^{-i\omega_0 t/2}$.

Finally, the squeeze operator is given by an exponential function of the creation and annihilation operators squared:
\begin{gather}\label{A06}
    S(w) = \exp\left\{\frac{1}{2}\left[w^*a^2-w\left(a^\dagger\right)^2\right]\right\},
\end{gather}
where $w$ is a complex number. The main properties of the rotation, displacement, and squeeze operators are reported in Appendix \ref{app:a}.

\subsubsection{Definition and transformation of the squeezed coherent states\label{subsec1}}

The relevant vibrational states are here represented by the so-called {\it squeezed coherent states}. These are obtained by sequentially applying a squeeze and a displacement operator to the vacuum state \cite{Gerry04a,Scully97a}
\begin{gather}
    |\alpha,z\rangle \equiv D(\alpha)\,S(z)\,|0\rangle .
\end{gather}

By exploiting the commutation relations between the rotation, squeeze, and displacement operators, one can show that their application to a squeezed coherent state gives rise to another squeezed coherent state (see Appendix \ref{app:a}). 
In particular, the application of the rotation operator to a squeezed coherent state modifies it as follows:
\begin{gather}\label{eq:44}
    R(\phi)\, (e^{i\zeta} | \alpha,z \rangle) = e^{i\zeta'} | \alpha',z' \rangle ,
\end{gather}
where the relation between the initial and final states is given by:
\begin{gather}
\alpha' = \alpha\, e^{i\phi},\ 
z' = z\, e^{2i\phi},\, \zeta'=\zeta .
\end{gather}

The application of the displacement operator modifies a squeezed coherent state according to the equation: 
\begin{gather}
D(\beta)\, (e^{i\zeta}| \alpha,z \rangle) = e^{i\zeta'} | \alpha',z' \rangle , 
\end{gather}
where initial and final states are related by:
\begin{gather}\label{eq:43}
\alpha' = \alpha + \beta,\ 
z' = z,\  
\zeta'=\zeta-i(\beta\alpha^* - \beta^*\alpha)/2.
\end{gather}

Finally, the application of the squeeze operator to a squeezed coherent state modifies it as follows: 
\begin{gather}
S(w)\, (e^{i\zeta}| \alpha,z \rangle ) = e^{i\zeta'} | \alpha',z' \rangle ,   
\end{gather}
where the relation between initial and final states is defined by the equations: 
\begin{gather}
\alpha' = \alpha\cosh |w| -\alpha^* e^{i\theta_w} \sinh |w|,\nonumber\\
e^{i\zeta'} = e^{i\zeta}\, \left( \frac{1+t_w t_z^*}{|1+t_w t_z^*|} \right)^{1/2} .
\label{eq:41}
\end{gather}
Here, $z=|z|e^{i\theta_z}$, $t_z$ is given by the expression
\begin{gather}\label{eq:z1}
t_z \equiv e^{i\theta_z} \tanh |z|,
\end{gather}
and an analogous relation applies to $w$ and $t_w$.
Consistently with the above definitions, the parameter $z'$ is given by the equation
\begin{gather}\label{eq:42}
t_{z'} = \frac{t_z+t_w}{1+t_z\,t_w^*},
\end{gather}
through the relations $|z'|={\rm atanh}\, |t_{z'}|$ and $e^{i\theta_{z'}}=t_{z'}/|t_{z'}|$.

\subsection{Reduction of the time evolution to squeezing, displacements and rotations\label{subsec:2}}

The vibrational time evolution operator corresponding to each time interval $t_k$ can be decomposed into the product of rotation, displacement, and squeeze operators \cite{Doktorov77a,Xin89a}. One possible decomposition, which can be generalized to the multimode case, is given by the expression
\begin{gather}
    U_k (t_k) \equiv e^{-i H_{v,\lambda_k} t_k/\hbar} = e^{-\omega_{\lambda_k} t_k/2}\,S(x_{\lambda_k}\!-\!x_0) \nonumber\\ D(-\delta_{\lambda_k})\, R(-\omega_{\lambda_k}t_k)\, D(\delta_{\lambda_k})\,S(x_0\!-\!x_{\lambda_k}).\label{eq:30}
\end{gather}
Here $x_{\lambda}\equiv\frac{1}{2}\ln\omega_{\lambda}$, {\color{black} with adimensional $\omega_\lambda$, obtained by dividing each angular frequency by a common reference value.} As a result, and in view of the equations reported in the previous Subsection, a squeezed coherent state whose dynamics is generated by an Hamiltonian $H_{v,\lambda}$ evolves into another squeezed coherent state, and can thus be identified by the two complex numbers $\alpha$ and $z$, plus the global phase $\zeta$. 

\subsubsection{Stepwise derivation of the final vibrational states\label{subsecX}}

The time evolution of the {\color{black} vibrational} state is generated by a Hamiltonian that is constant during each time interval $t_k$, and changes from one interval to the other [Eq. (\ref{eq:51})].
We call $e^{i\zeta_{k,i}}|\alpha_{k,i},z_{k,i}\rangle $ and $e^{i\zeta_{k,f}}|\alpha_{k,f},z_{k,f}\rangle$ the squeezed coherent state respectively at the beginning and at the end of $t_k$. 
The final state for each time interval coincides with the initial state for the following one: 
\begin{gather}
e^{i\zeta_{k,i}}|\alpha_{k,i},z_{k,i}\rangle \equiv  e^{i\zeta_{k-1,f}}|\alpha_{k-1,f},z_{k-1,f}\rangle .
\end{gather}
In order to derive the vibrational state at the end of the time interval, one can first reduce the time evolution operator $e^{-i H_{v,\lambda_k} t_k/\hbar}$ to the product of squeeze, displacement, and rotation operators [Eq. (\ref{eq:30})], and then sequentially apply such operators to the vibrational state. The resulting intermediate states, referred to as $e^{i\zeta_{k,l}}|\alpha_{k,l},z_{k,l}\rangle$ (with $l=1,\dots,5$), are derived as follows:
\begin{enumerate}
    \item The first intermediate state is obtained by applying the squeeze operator, and thus of Eqs. (\ref{eq:41}-\ref{eq:42}), with $w = \frac{1}{2}\ln(\omega_0/\omega_{\lambda_k})$.
    The states before and after such application are specified by $(\alpha,z,\zeta) = (\alpha_{k,i},z_{k,i},\zeta_{k,i})$ and
    $(\alpha',z',\zeta') = (\alpha_{k,1},z_{k,1},\zeta_{k,1})$.
    \item The second intermediate state is obtained by applying a displacement operator, and thus of Eq. (\ref{eq:43}), with $\beta = \delta_{\lambda_k}$. The states before and after such application are specified by 
    $(\alpha,z,\zeta) = (\alpha_{k,1},z_{k,1},\zeta_{k,1})$ and
    $(\alpha',z',\zeta') = (\alpha_{k,2},z_{k,2},\zeta_{k,2})$.
    \item The third intermediate state results from the application of a rotation operator [Eq. (\ref{eq:44})], with $\phi = -\omega_{\lambda_k} t_k$. The states before and after such application are specified by  
    $(\alpha,z,\zeta) = (\alpha_{k,2},z_{k,2},\zeta_{k,2})$ and
    $(\alpha',z',\zeta') = (\alpha_{k,3},z_{k,3},\zeta_{k,3})$.
    \item The fourth intermediate state is obtained by applying a displacement operator [Eq. (\ref{eq:43})], with $\beta = -\delta_{\lambda_k}$. The states before and after such application are specified by 
    $(\alpha,z,\zeta) = (\alpha_{k,3},z_{k,3},\zeta_{k,3})$ and
    $(\alpha',z',\zeta') = (\alpha_{k,4},z_{k,4},\zeta_{k,4})$.
    \item The fifth state squeezed coherent state results from the application of a squeeze operators, and thus of Eqs. (\ref{eq:41}-\ref{eq:42}), with $w = \frac{1}{2}\ln(\omega_{\lambda_k}/\omega_0)$. The states before and after such application are specified by 
    $(\alpha,z,\zeta) = (\alpha_{k,4},z_{k,4},\zeta_{k,4})$, and
    $(\alpha',z',\zeta') = (\alpha_{k,5},z_{k,5},\zeta_{k,5})$.
\end{enumerate}
The vibrational state at the end of the waiting time $t_k$ is then given by $\alpha_{k,f}=\alpha_{k,5}$, $z_{k,f}=z_{k,5}$, and $\zeta_{k,f}=\zeta_{k,5}-\omega_{\lambda_k} t_k/2$. We stress that the $e^{i\zeta_{k,l}}|\alpha_{k,l},z_{k,l}\rangle$ are fictitious intermediate states: the vibrational state does not go through these states during its time evolution, but it does coincide with $e^{i\zeta_{k,i}}|\alpha_{k,i},z_{k,i}\rangle $ and $e^{i\zeta_{k,f}}|\alpha_{k,f},z_{k,f}\rangle$, respectively at the beginning and at the end of the waiting time $t_k$. 

\subsection{Vibrational response functions}

The vibrational component of the $M$-th order response function is given by the overlap between the final vibrational state $e^{i\zeta}|\alpha,z\rangle$, corresponding to the state at the end of the $M$-th waiting time, and the initial vibrational state, coinciding with the ground state of $H_{v,0}$:
\begin{gather}
    R^{(v,M)}_{\Lambda} ({\bf t}) = \langle \psi_{bra} | \psi_{ket,\Lambda} \rangle =  
    e^{i\zeta} \langle 0 | \alpha, z \rangle = \nonumber\\ 
    \frac{e^{i\zeta}}{\sqrt{\cosh |z|}} \exp\left\{-\frac{1}{2}\left[|\alpha|^2-t_z (\alpha^*)^2\right]\right\},\label{eq:rf1}
\end{gather}
where $t_z$ is given by Eq. (\ref{eq:z1}), and the squeezed coherent state is given by $\alpha=\alpha_{M,f}$, $z=z_{M,f}$, and $\zeta=\zeta_{M,f}+\omega_{0}\sum_{k=1}^M t_k/2$.

If the initial state of the vibrational mode does not coincide with the ground state of $H_{v,0}$, but with a generic squeezed coherent state, $|\alpha_0,z_0\rangle$, the above procedure for the derivation of the vibrational response function has to be slightly modified, by adding to the sequence a squeeze, a rotation, and a displacement operator: 
\begin{gather}\label{xyz}
    R^{(v,M)}_{\Lambda} ({\bf t}) 
    \!=\! \langle \psi_{bra} | \psi_{ket,\Lambda} \rangle 
    \!=\! e^{i\phi_0/2}\, \langle 0 | \psi_{ket,\Lambda}' \rangle ,
\end{gather}
where $\phi_0=\omega_0\sum_{j=1}^M t_j$ and the effective ket state reads:
\begin{gather}\label{eq:51x}
    |\psi_{ket,\Lambda}'\rangle = S(-z_0)\, D(-\alpha_0)\, R(\phi_0) \nonumber\\ \left(e^{-i H_{v,\lambda_M}t_M/\hbar} \dots e^{-i H_{v,\lambda_1}t_1/\hbar}\right) | \alpha_0 , z_0 \rangle .
\end{gather}

\subsection{Application to some representative case\label{subsec:1d}}

For simplicity, we apply the above approach to the calculation of the first- and third-order response functions of a two-level system ($N_e=2$). The double-sided Feynman diagrams that are relevant for the first- and third-order response functions and where all the interactions with the field occur on the left (ket) are shown in Fig. \ref{fig:0}.

\subsubsection{First-order response functions}

\begin{figure}[h]
\centering
\includegraphics[width=0.35\textwidth]{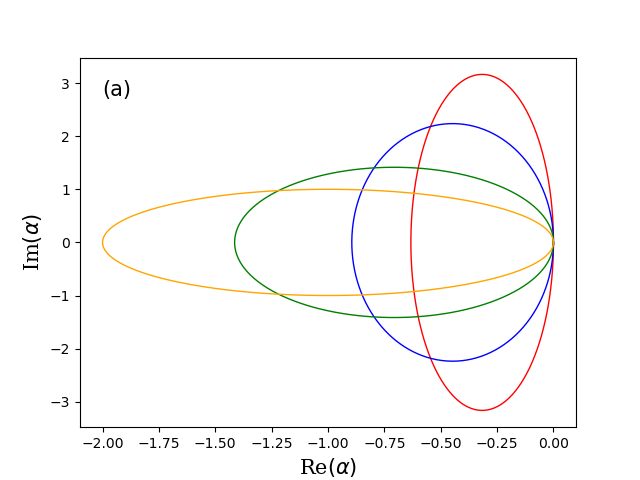}
\includegraphics[width=0.35\textwidth]{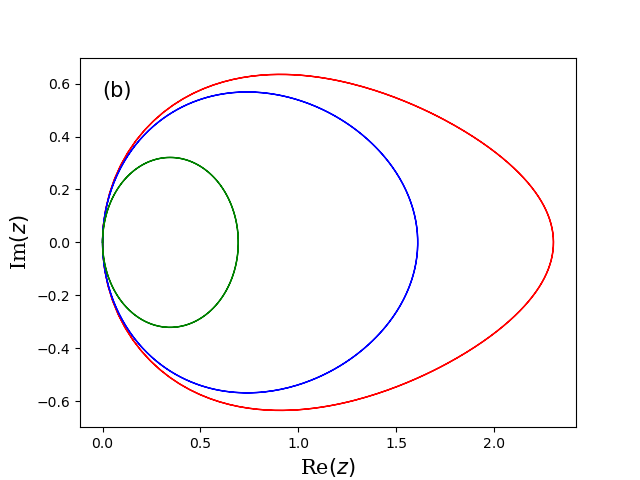}
\includegraphics[width=0.35\textwidth]{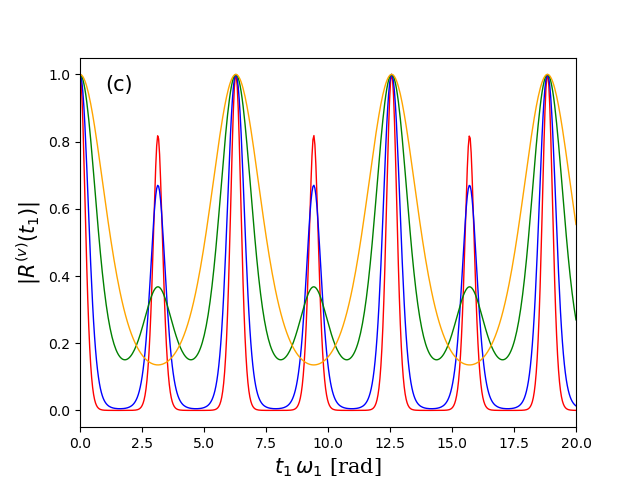}
\caption{Trajectory described by the squeezed coherent state, defined by the values of the complex numbers (a) $\alpha$ and (b) $z$. {\color{black} Please note that for visualization purposes, the scales of the two axes are different.} (c) Time dependence of the vibrational component of the linear response function ($M=1$). The curves in the three panels are obtained for a same value of the displacement $\delta_1=1$, and for different values of the ratio {\color{black} $\omega_1/\omega_0$} between ground- and excited-state vibrational frequencies: {\color{black} $10$ (red), $5$ (blue), $2$ (green), $1$ (orange)}.} 
\label{fig:1}
\end{figure}

In the linear case ($M=1$), the electronic component of the response function is given by \cite{Hamm11a}
\begin{gather}\label{eq:erf1}
    R^{(e,1)}_{\Lambda}(t_1) = i |\mu_{01}|^2\, e^{-i\epsilon_1t_1/\hbar}\,e^{-(\gamma+\Gamma/2)t_1},
\end{gather}
where $\epsilon_0=0$, {\color{black} $\Lambda=(1)$,} and $\gamma$ ($\Gamma$) is the dephasing (relaxation) rate. {\color{black} The effective inclusion of dephasing and relaxation into a closed-system approach is based on some specific assumption: first, that dephasing results in the (exponential) decay of the coherences generated by the electric field, and doesn't give rise to a transfer between one coherence and the other \cite{CT}; second, that the pathways generated by the relaxation process through population transfer can be neglected.}

The vibrational response function is obtained by applying the equations reported above in Subsections \ref{subsec:1}-\ref{subsec:2}, and specifically for $M=1$ and $\lambda_1=1$. In fact, these equations provide the time evolution of the vibrational state, {\it i.e.} the dependence on $t_1$ of the state 
$|\psi_{ket,\Lambda}\rangle=e^{i\zeta}|\alpha,z\rangle$, whose overlap with the initial state $|0_0\rangle$ gives the vibrational response function [Eq. (\ref{eq:52})]. 
The global phase $\zeta$ is specific of each electronic pathway $p_e$, and thus determines the algebraic sum of contributions resulting from different pathways. While focusing on a single pathway, the discussion can be limited to the time evolution of the two complex numbers, $\alpha$ and $z$.  

In Fig. \ref{fig:1}(a), we plot the trajectory described by $\alpha$ in the plane $[\alpha_r\equiv$Re$(\alpha),\alpha_i\equiv$Im$(\alpha)]$, as a function of the waiting time $t_1$. We note that the real and imaginary parts of $\alpha$ coincide with the expectation values of the quadrature operators $X_1=(a+a^\dagger)/2$ and $X_2=(a-a^\dagger)/2i$, respectively. In the particular case where the electronic excitation implies a displacement of the oscillator, but not a change in the vibrational frequency ($\omega_1=\omega_0$), $\alpha$ describes a circle with radius $|\delta_1|$, centered in $(-\delta_1,0)$ \cite{Quintela2022a}. {\color{black} For increasing values of $\omega_1/\omega_0$}, the trajectories described by the parameter $\alpha$ are increasingly elongated along $\alpha_i$. As results from Eq. (\ref{eq:30}), the time dependence of $\alpha$ is periodic, with a periodicity that  depends on the excited-state vibrational frequency through the relation $\omega_1 t_1=2 k\pi$. However, but the trajectory described in the $(\alpha_r,\alpha_i)$ plane only depends on the displacement and on the ratio between the frequencies (the same applies to $z$, see below).
The squeezing parameter $z$ also describes a closed trajectory in the $(z_r,z_i)$ plane, and specifically in the region $z_r\ge 0$ [Fig. \ref{fig:1}(b)]. In particular, for $\omega_1=\omega_0$ no squeezing occurs, and $z$ remains identically zero. In the presence of an electronic-state dependence of the vibrational frequency, and specifically for $\omega_1\neq\omega_0$, the trajectories described by $z$ range from $z_r=0$ to $z_r=$ln$(\omega_1/\omega_0)$, the latter corresponding to a maximum squeezing along the $X_1$ direction. 

The modulus of the vibrational response function corresponds to that of the overlap between $|\alpha,z\rangle$ and $|0\rangle$ [Eq. (\ref{eq:rf1})]. Such overlap is always maximum for $t_1\omega_1=2 k\pi$, where the final and initial vibrational states coincide. In the absence of squeezing, $R^{(v,1)_{p_e}}(t_1)$ presents minima for $t_1\omega_1=(2 n+1)\pi$, where $|\alpha|$ is maximal and is given by $2|\delta_1|$. For decreasing values of $\omega_1/\omega_0$, however, local maxima of increasing height appear at $t_1\omega_1=(2 n+1)\pi$, due to the decreasing distance between the origin of the $(\alpha_r,\alpha_i)$ plane and the second intersection with the $\alpha_r$ axis [Fig. \ref{fig:1}(a)]. For $\omega_0>\omega_1$ (not reported here), the trajectory of $\alpha$ is elongated along $\alpha_r$, that of $z$ is contained in the half plane $z_r<0$, and the local maxima in the modulus of the vibrational response function are absent. 
All this shows a clear connection between the electronic-state dependent change of the vibrational frequency, the trajectory of the squeezed coherent state, and features observed in the response function.

\subsubsection{Third-order response functions}

For the third-order term ($M=3$), we specifically consider the non-rephasing part of the ground-state bleaching contribution [$\lambda_1=\lambda_3=1$ and $\lambda_2=0$, Fig. \ref{fig:0}(b)].
The electronic part of the response function is given by \cite{Hamm11a}
\begin{gather}\label{eq:erf3}
    R_\Lambda^{(e,3)}({\bf t}) = -i |\mu_{01}|^4\, e^{-i\epsilon_1(t_1+t_3)/\hbar}\,e^{-(\gamma+\Gamma/2)(t_1+t_3)}
\end{gather}
where $\epsilon_0=0$, {\color{black}$\Lambda=(1,0,1)$,} and $\gamma$ ($\Gamma$) is the dephasing (relaxation) rate.

\begin{figure}[h]
\centering
\includegraphics[width=0.23\textwidth]{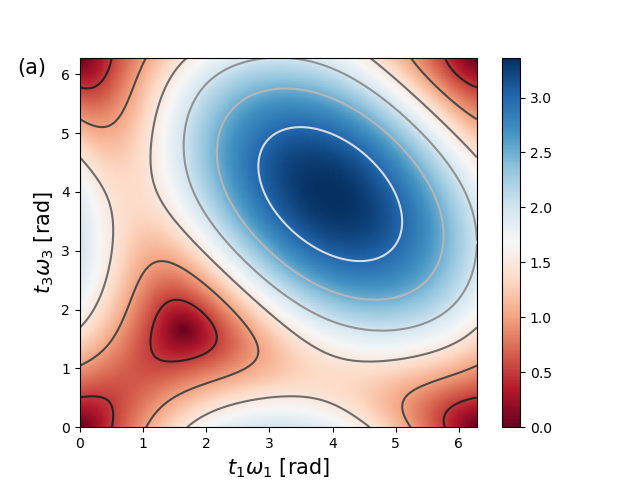}
\includegraphics[width=0.23\textwidth]{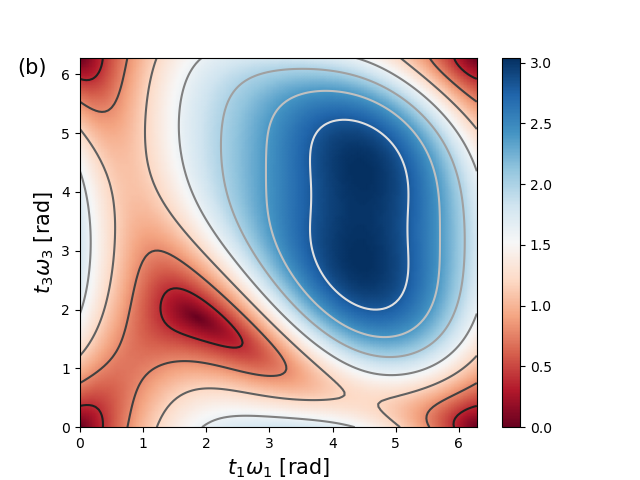}
\includegraphics[width=0.23\textwidth]{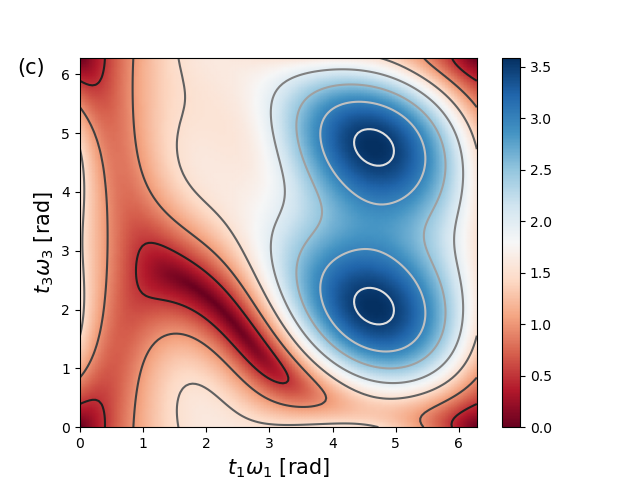}
\includegraphics[width=0.23\textwidth]{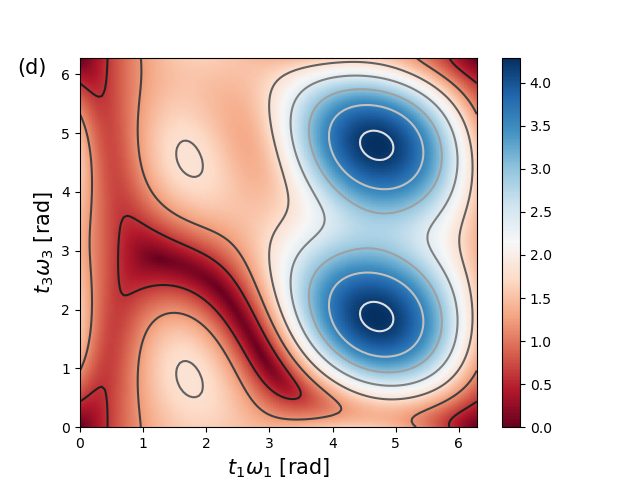}
\includegraphics[width=0.23\textwidth]{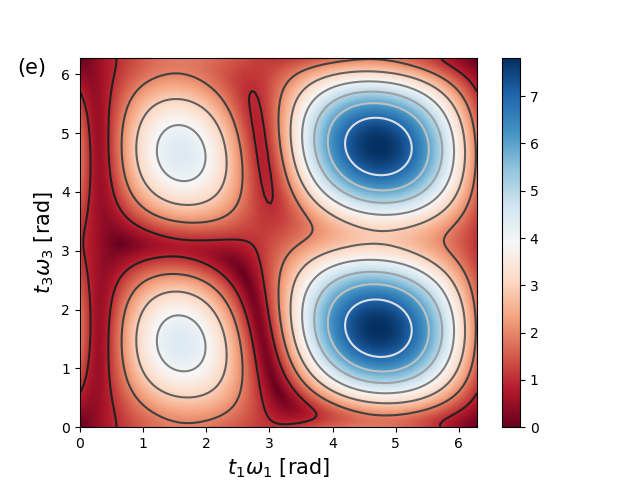}
\includegraphics[width=0.23\textwidth]{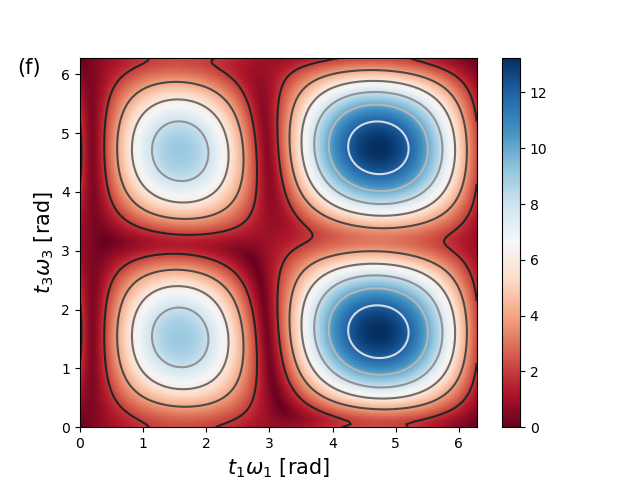}
\caption{Absolute value of $\alpha$ in the case $M=3$, as a function of the waiting times $t_1$ and $t_3$, and for $t_2\omega_0=1.5\,$rad. Different panels correspond to different values of {\color{black} $\omega_1/\omega_0$, and specifically: (a) $1$, (b) $1.125$, (c) $5/3$, (d) $2$, (e) $2.5$, (f) $5$. In all cases, the displacement is given by $\delta_1=1$.}} 
\label{fig:S1}
\end{figure}

The parameters $\alpha$ and $z$ are themselves functions of the three waiting times. For fixed values of two times, they describe closed trajectories as a function of the third time, similar to those shown above for the case of linear response functions.  
The dependence of $|\alpha|$ on the first and third waiting times, for a fixed value of $t_2$ and decreasing values of {\color{black}$\omega_1/\omega_0$}, is reported in Fig. \ref{fig:S1} [panels (a) to (f)]. The displayed time ranges are limited to $0\le t_k\omega_k\le 2\pi$ because $\alpha$ is a periodic function of the waiting times (and so is $z$), being its dependence on $t_k$ entirely contained in that of $R(-\omega_{\lambda_k} t_k)$ [Eq. (\ref{eq:30})], which coincides with $R(-\omega_{\lambda_k} t_k+2\pi)$. The comparison between the different panels shows qualitative changes as a function of the ratio between the ground- and excited-state frequencies: in particular, a single maximum shows up for $\omega_0=\omega_1$ (a), while four local maxima appear for $\omega_1\ge 2.5\,\omega_0$ (e,f). Qualitative changes with $\omega_0=\omega_1$ also occur to the local minima, corresponding to the regions in the $(t_1,t_3)$ plane where the final vibrational state is closer to the initial one.  

\begin{figure}[h]
\centering
\includegraphics[width=0.23\textwidth]{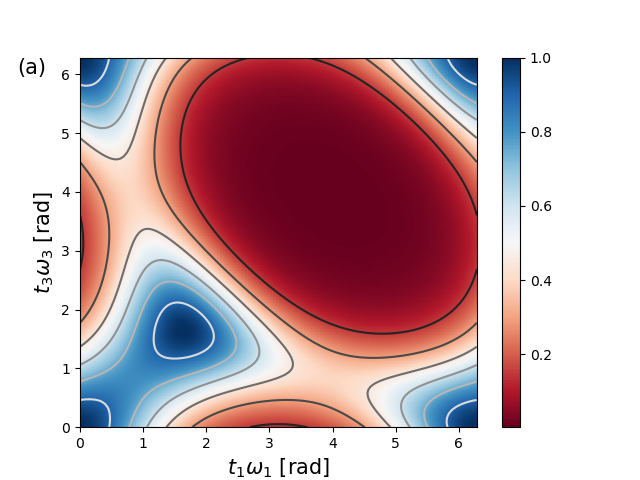}
\includegraphics[width=0.23\textwidth]{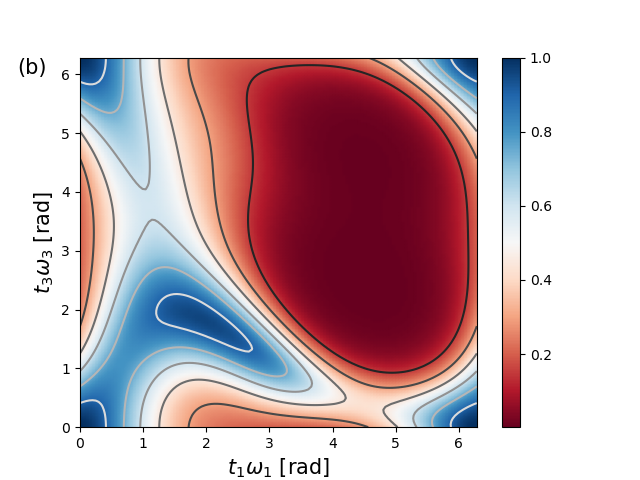}
\includegraphics[width=0.23\textwidth]{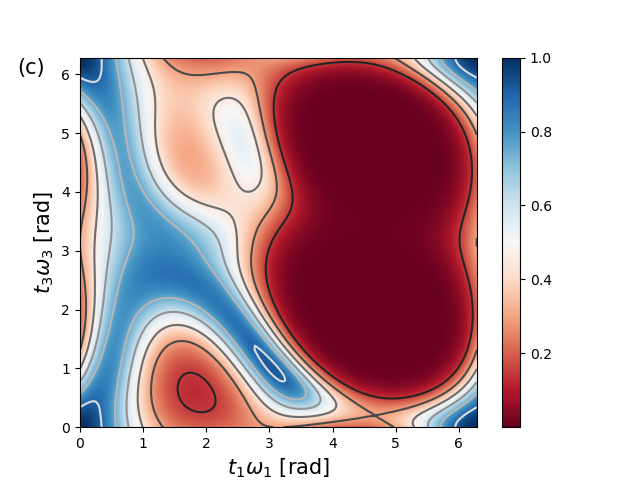}
\includegraphics[width=0.23\textwidth]{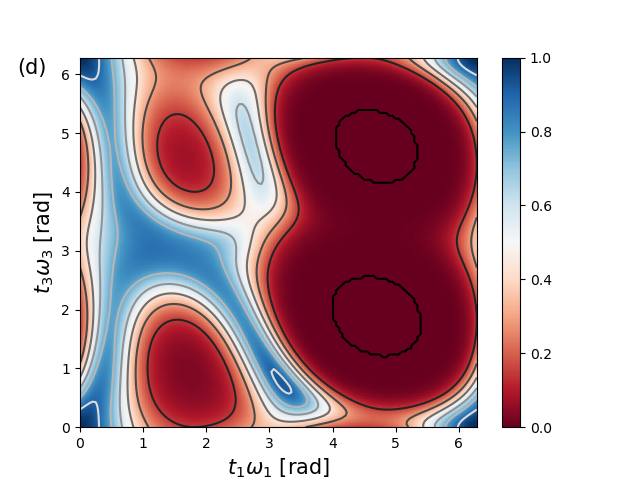}
\includegraphics[width=0.23\textwidth]{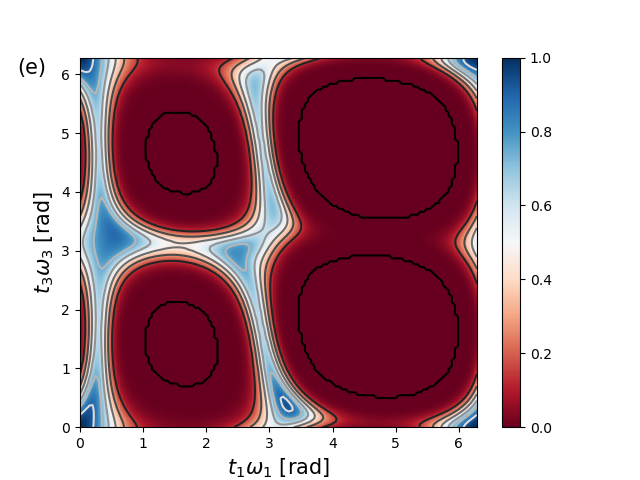}
\includegraphics[width=0.23\textwidth]{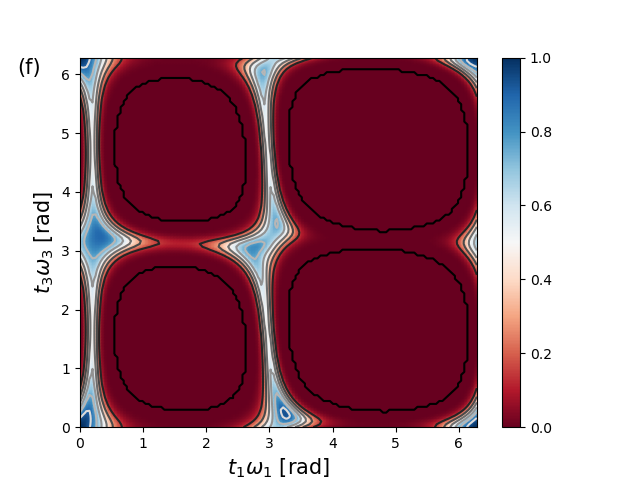}
\caption{Absolute value of the third-order vibrational response function $R^{(v)}$ in the case $M=3$, as a function of the waiting times $t_1$ and $t_3$, and for $t_2\omega_0=1.5\,$rad. Different panels correspond to different values of {\color{black}$\omega_1/\omega_0$, and specifically: (a) $1$, (b) $1.125$, (c) $5/3$, (d) $2$, (e) $2.5$, (f) $5$. In all cases, the displacement is given by $\delta_1=1$.}}
\label{fig:2}
\end{figure}

Besides these specific features, which can vary for different values of $t_2$, we would like to stress the clear correlation between the behavior of $|\alpha|$ and that of the response function (Fig. \ref{fig:2}). In fact, this displays minima (maxima) in the same regions where $|\alpha|$ displays its maxima (minima), and undergoes the same changes as a function of the ratio {\color{black} $\omega_1/\omega_0$ between the vibrational frequencies corresponding to the excited and ground electronic states}. Such correlation results from the identification of the response function with the overlap between the initial and the final vibrational states, $|0,0\rangle$ and $e^{i\zeta} |\alpha,z\rangle$, dominated by the exponential dependence on $\alpha$ [Eq. (\ref{eq:rf1})]. Though complicated by the presence of multiple waiting times, also these trends clearly show the connection between the vibrational response function and the dynamics of the vibrational wave packet.

\section{Multiple vibrational modes with electronic-state dependent frequency}

In the presence of multiple modes, a change in the electronic state can also result in a rotation between the vibrational modes. The approach outlined so far can be generalized to deal with the multi-mode case, as shown in the following.

\subsection{Definition of the model}

The model system consists of $N_e$ electronic levels and $N$ vibrational modes. These are specifically given by a set of harmonic oscillators, whose frequency and origin depend on the electronic state. The corresponding Hamiltonian reads:
\begin{align}
  H = \sum_{\lambda=0}^{N_e} |\lambda\rangle\langle\lambda | \otimes \left( \epsilon_\lambda + H_{v,\lambda} \right),
\end{align}
where the vibrational Hamiltonians $H_{v,\lambda}$ are given by:
\begin{gather}\label{eq:mmvh}
    H_{v,\lambda} 
    \!=\! \sum_{j=1}^{N} \omega_{\lambda,j} (a_{\lambda,j}^\dagger\,a_{\lambda,j}+1/2).
\end{gather}
The modes corresponding to the electronic states $\lambda$ and $0$ (ground state) are assumed to be linearly related, and more specifically by the Duschinsky rotations \cite{Doktorov77a,Huh15a}:
\begin{gather}
    {\bf q}_\lambda = U_\lambda \, {\bf q}_0 + {\bf d}_\lambda,
\end{gather}
being $U_\lambda$ the $N \times N$ orthogonal Duschinsky rotation matrix and ${\bf d}_\lambda$ the displacement vector. The components of the vector ${\bf q}_\lambda$ are the position operators $q_{\lambda,j}$ ($j=1,\dots,N$), related to the creation and annihilation operators by the equations
$q_{\lambda,j}=\sqrt{\hbar/(2\omega_{\lambda,j})}(a_{\lambda,j}+a_{\lambda,j}^\dagger)$. Correspondingly, the displacements of the position and annihilation (creation) operators are related by the equations $d_{\lambda,j}=\sqrt{2\hbar/\omega_{\lambda,j}}\,\delta_{\lambda,j}$.

As in the single-mode case, the time evolution of the vibrational state is induced by an Hamiltonian that is piecewise constant, and coincides at each time interval $t_k$ with a $H_{v,\lambda_k}$, being $\lambda_k$ an electronic state. As a result Eqs. (\ref{eq:51}-\ref{eq:52}) still apply, but $|\psi_{bra}\rangle$ and $|\psi_{ket,\Lambda}\rangle$ are now multi-mode vibrational states. 

\subsection{Rotation, displacement, and squeeze operators\label{subsec:1p}}

In analogy with the single-mode case, we exploit the fact that a multi-mode squeezed coherent state evolves, under the effect of the Hamiltonians $H_{v,\lambda}$, into another squeezed coherent state. This again results from the fact that the time evolution operator can be decomposed into the product of rotation, squeeze, and displacement operators \cite{Doktorov77a}, and that each of these operators preserves the squeezed coherent character of the vibrational state.

\subsubsection{Definition of the operators}

The rotation, displacement, and squeeze operators can be generalized to the multimode case \cite{Xin90a}.
The rotation operator can be expressed as an exponential function of the creation and annihilation operators, according to the equation
\begin{gather}
    R_N(\Phi) = \exp\left(i \tilde a^\dagger\,\Phi\, a\right) .
\end{gather}
Here $\Phi$ is a Hermitian matrix and $\tilde a = (a_1,\dots,a_{N_e})$  and $\tilde a^\dagger = (a_1^\dagger,\dots,a_{N_e}^\dagger)$ {\color{black} (the tilde represents the transposition of the $N_e \times 1$ matricial operators $a$ and $a^\dagger$)}. The time evolution operator generated by $H_{v,0}$ corresponds to a rotation with a diagonal matrix $\Phi$ and nonzero elements $\Phi_{jj}=-\omega_{0,j}\,t$, multiplied by a phase factor $\prod_{j=1}^{N} e^{-i\omega_{0,j}t/2}$. 

The displacement operator can also be written as an exponential function of the creation and annihilation operators:
\begin{gather}\label{A03p}
    D_N(B) = \exp\left(\tilde B a^\dagger - B^\dagger a\right) ,
\end{gather}
where $\tilde B=(\beta_1,\dots,\beta_P)$ {\color{black} (the tilde and the dagger applied to a numerical matrix denote its simple and conjugate transpositions, respectively)}. One can easily verify that, unlike the rotation and squeeze operator (see below), the multimode displacement operator can always be written as the product of $N$ single-mode displacement operators: $D_N(B) = \prod_{j=1}^{N} D(\beta_j) $. 

Finally, the squeeze operator is given by an exponential function of a quadratic function of the creation and annihilation operators:
\begin{gather}\label{A06p}
    S_N(Z) = \exp\left[\frac{1}{2}\left(\tilde a Z^\dagger a-\tilde a^\dagger Z a^\dagger\right)\right],
\end{gather}
where $Z$ is a complex symmetric matrix. The main properties of the multi-mode rotation, squeeze, and displacement operators are recalled in Appendix \ref{app:b}. 

\subsubsection{Definition and transformation of the squeezed coherent states}

In analogy with the single-mode case, we identify the multi-mode squeezed coherent state with
\begin{gather}
    |A,Z\rangle = D_N(A)\,S_N(Z)\,|0\rangle ,
\end{gather}
where $|0\rangle$ is the multi-mode vacuum state. 
If applied to a multi-mode squeezed coherent state, the operators $S_N(Z)$,  $D_N(B)$, and $R_N(\Phi)$ generate another squeezed coherent state (see Appendix \ref{app:b} for further details).

The application of the rotation operator to a squeezed coherent state induces the following transformation:
\begin{gather}\label{eq:roi}
    R_N(\Phi)\, (e^{i\zeta}| A, Z \rangle) = e^{i\zeta'}| A',Z' \rangle  
\end{gather}
where the parameters defining the initial and final states are related by 
\begin{gather}\label{eq:rof}
A' = e^{i\Phi}\,A,\ 
Z' = e^{i\Phi}\, Z\, e^{i\tilde\Phi},\,\zeta'=\zeta 
\end{gather}
and $\tilde\Phi$ is the transpose of the matrix $\Phi$.

The application of the displacement operator modifies a squeezed coherent state according to the equations
\begin{gather}\label{eq:doi}
D_N(B)\, (e^{i\zeta}| A,Z \rangle) = e^{i\zeta'} | A',Z' \rangle ,
\end{gather}
where the parameters defining the initial and final states are related by 
\begin{gather}\label{eq:dof}
A' = A + B,\ 
Z' = Z, \ 
\zeta'=\zeta-i(A^\dagger B - B^\dagger A)/2.
\end{gather}

The application of the squeeze operator modifies a squeezed coherent state according to the equations
\begin{gather}\label{eq:soi}
S_N(W)\, (e^{i\zeta}| A,Z \rangle) = e^{i\zeta'} | A',Z' \rangle    
\end{gather}
where the transformed state is defined by 
\begin{gather}
A' = \cosh |W|\,A - \sinh |W|\, e^{i\Theta_W}\,A^*,
\end{gather}
where $W=|W|e^{i\Theta_W}$ is the polar decomposition of the $N\times N$ matrix $W$. 
The matrix $Z'$ is determined by the equation
\begin{gather}
T_{Z'}=S_W^{-1} (T_W+T_Z) (I+T_W^\dagger T_Z)^{-1} \tilde S_W,
\end{gather}
where $T_W\equiv \tanh |W| e^{i\Theta_W}$ (analogously for $Z$ and $Z'$) and $S_W\equiv {\rm sech} |W|$.
Finally, the phase factor results from
\begin{gather}
\zeta'=\zeta+\frac{1}{2} {\rm Tr} (\Phi) .
\end{gather}
with
\begin{gather}\label{eq:sof}
e^{i\Phi} = S_{Z'}^{-1} S_W (I+T_Z T_W^\dagger)^{-1} S_Z .
\end{gather}

\subsection{Reduction of the time evolution to squeezing, displacements and rotations\label{subsec:2p}}

In analogy to what has been done in the single-mode case, the vibrational time evolution operator $e^{-i H_{v,\lambda_k} t_k/\hbar}$ can be written as the product of displacement, squeeze and rotation operators, according to the expression \cite{Doktorov77a}:
\begin{gather}
U_k(t_k) = e^{-\sum_{j=1}^{N}\omega_{\lambda_k,j} t/2} S_N^\dagger(X_0)\,R_N^\dagger(\Phi_{\lambda_k})\,S_N(X_{\lambda_k}) \nonumber\\ D_N^\dagger(\Delta_{\lambda_k})\, R_N(\Phi_{\lambda_k}') D_N(\Delta_{\lambda_k})\,S_N^\dagger(X_{\lambda_k})\,R_N(\Phi_{\lambda_k})\,S_N(X_0) . \label{eq:50}
\end{gather}
Here the squeeze operators are defined by the matrix 
\begin{gather}
X_\lambda=\frac{1}{2}\, {\rm diag} (\ln\omega_{\lambda,1},\dots,\ln\omega_{\lambda,N})    
\end{gather}
and the displacement operators are defined by the vector 
\begin{gather}
\tilde\Delta_\lambda = (\delta_{\lambda,1},\dots,\delta_{\lambda,N}) .
\end{gather}
As to the rotation operators, the first and last ones are defined by the matrix
\begin{gather}
\Phi_{\lambda} = -i\ln U_\lambda   
\end{gather}
while the second rotation is defined by the diagonal matrix
\begin{gather}
\Phi_{\lambda}' = -t_k \, {\rm diag} (\omega_{\lambda,1},\dots,\omega_{\lambda,N})  .
\end{gather}
If there is no mixing between the modes ($\Phi_{\lambda_k}=0$), $U_k(t_k)$ is reduced to the product of $N$ single-mode time evolution operators, whose expression is given in Eq. (\ref{eq:30}).  

\subsubsection{Stepwise derivation of the final vibrational states}

As in the single-mode case, the time evolution of the {\color{black} vibrational} state is generated by a Hamiltonian that is constant during each time interval $t_k$, and changes from one interval to the other [Eq. (\ref{eq:51})].
We call $e^{i\zeta_{k,i}}|A_{k,i},Z_{k,i}\rangle $ and $e^{i\zeta_{k,f}}|A_{k,f},Z_{k,f}\rangle$ the squeezed coherent state respectively at the beginning and at the end of the time interval $t_k$. 
The final state for each time interval coincides with the initial state for the following one: 
\begin{gather}
e^{i\zeta_{k,i}}|A_{k,i},Z_{k,i}\rangle \equiv  e^{i\zeta_{k-1,f}}|A_{k-1,f},Z_{k-1,f}\rangle 
\end{gather}
\begin{figure}[h]
\centering
\includegraphics[width=0.35\textwidth]{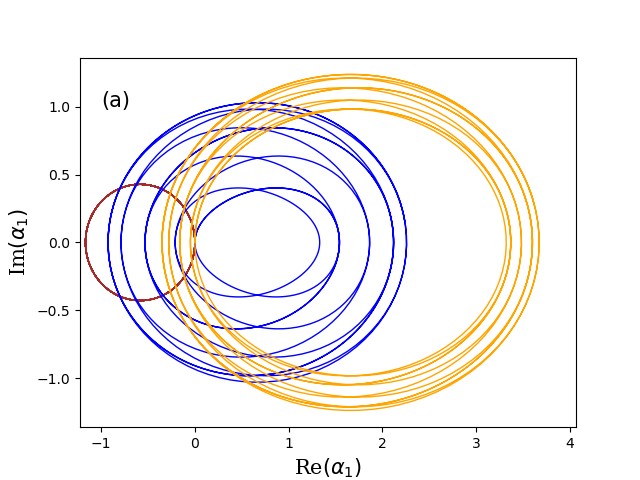}
\includegraphics[width=0.35\textwidth]{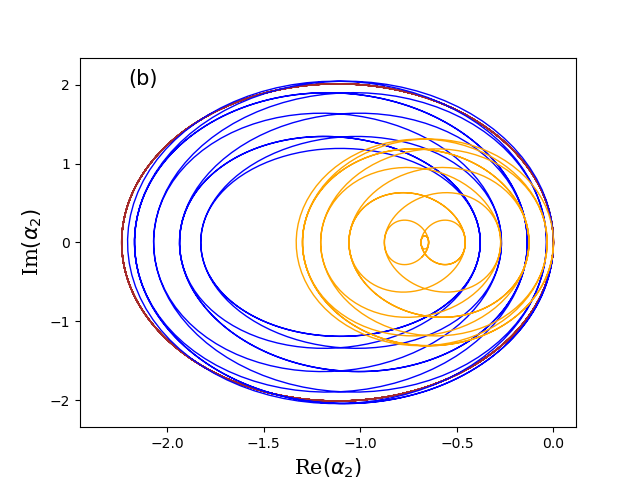}
\includegraphics[width=0.35\textwidth]{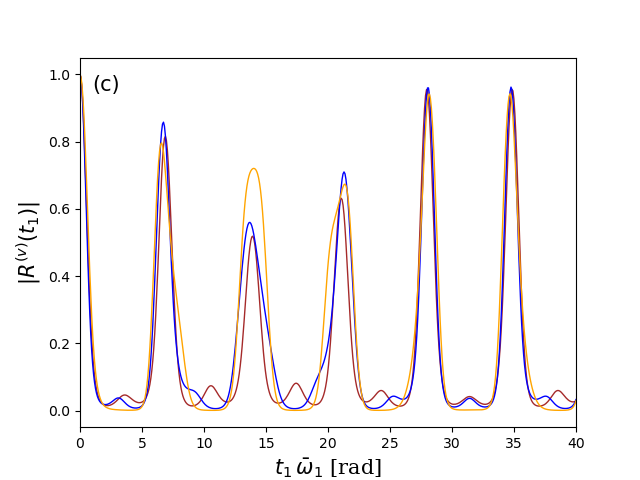}
\caption{Trajectory of the two-mode squeezed coherent state as a function of $t_1$, defined by the values of (a) $\alpha_1$ and (b) $\alpha_2$. {\color{black} Please note that for visualization purposes, the scales of the two axes are different.} (c) Time dependence of the linear response function ($N=1$), with $\bar\omega_1\equiv (\omega_{1,1}+\omega_{1,2})/2$. The curves in the three panels are obtained for $\delta_{1,1}=0.5$ and $\omega_{1,1}/\omega_{0,1}=0.73$ (first mode), $\delta_{1,2}=1.5$ and $\omega_{1,2}/\omega_{0,2}=1.8$ (second mode). Curves of different colors correspond to different values of the rotation angle $\varphi_1/\pi$, and specifically: $0$ (brown), $0.2$ (blue), $0.4$ (orange).}
\label{fig:3}
\end{figure}
In order to derive the vibrational state at the end of the time interval, one can first reduce the time evolution operator to the product of squeeze, displacement, and rotation operators [Eq. (\ref{eq:50})], and then sequentially apply such operators to the vibrational state. The resulting intermediate states, referred to as $e^{i\zeta_{k,l}}|A_{k,l},Z_{k,l}\rangle$, are given by:
\begin{enumerate}
    \item The first intermediate state is obtained by applying the squeeze operator, and thus Eqs. (\ref{eq:soi}-\ref{eq:sof}), with $W = X_0$.
    The states before and after such application are specified by
    $(A,Z,\zeta) = (A_{k,i},Z_{k,i},\zeta_{k,i})$ and
    $(A',Z',\zeta') = (A_{k,1},Z_{k,1},\zeta_{k,1})$. 
    \item The second intermediate state is obtained by applying a rotation operator, and thus Eqs. (\ref{eq:roi}-\ref{eq:rof}), with $\Phi = \Phi_{\lambda_k}$.
    The states before and after such application are specified by
    $(A,Z,\zeta) = (A_{k,1},Z_{k,1},\zeta_{k,1})$,
    $(A',Z',\zeta') = (A_{k,2},Z_{k,2},\zeta_{k,2})$.
    \item The third intermediate state is obtained by applying a second squeeze operator, and thus results from Eqs. (\ref{eq:soi}-\ref{eq:sof}) with $W = -X_{\lambda_k}$.
    The states before and after such application are specified by
    $(A,Z,\zeta) = (A_{k,2},Z_{k,2},\zeta_{k,2})$,
    $(A',Z',\zeta') = (A_{k,3},Z_{k,3},\zeta_{k,3})$.
    \item The fourth intermediate state is obtained by applying a displacement operator, whose effect on the squeezed coherent state is given by Eqs. (\ref{eq:doi}-\ref{eq:dof}), with $B = \Delta_{\lambda_k}$.
    The states before and after such application are specified by
    $(A,Z,\zeta) = (A_{k,3},Z_{k,3},\zeta_{k,3})$,
    $(A',Z',\zeta') = (A_{k,4},Z_{k,4},\zeta_{k,4})$.
    \item The fifth intermediate state results from the application of a rotation operator, whose effect is given by Eqs. (\ref{eq:roi}-\ref{eq:rof}), with $\Phi = \Phi_{\lambda_k}'$.
    The states before and after such application are specified by
    $(A,Z,\zeta) = (A_{k,4},Z_{k,4},\zeta_{k,4})$,
    $(A',Z',\zeta') = (A_{k,5},Z_{k,5},\zeta_{k,5})$.
    \item The sixth intermediate state is obtained by applying a displacement operator, and thus Eqs. (\ref{eq:doi}-\ref{eq:dof}), with $B = -\Delta_{\lambda_k}$.
    The states before and after such application are specified by
    $(A,Z,\zeta) = (A_{k,5},Z_{k,5},\zeta_{k,5})$,
    $(A',Z',\zeta') = (\alpha_{k,6},z_{k,6},\zeta_{k,6})$.
    \item The seventh intermediate state results from the application of a squeeze operator [Eqs. (\ref{eq:soi}-\ref{eq:sof})], with $W = X_{\lambda_k}$.
    The states before and after such application are specified by
    $(A,Z,\zeta) = (A_{k,6},Z_{k,6},\zeta_{k,6})$,
    $(A',Z',\zeta') = (A_{k,7},Z_{k,7},Z_{k,7})$.
    \item The eighth intermediate state is obtained from the application of a rotation operator, and thus from Eqs. (\ref{eq:roi}-\ref{eq:rof}), with $\Phi = -\Phi_{\lambda_k}$.
    The states before and after such application are specified by
    $(A,Z,\zeta) = (A_{k,7},Z_{k,7},\zeta_{k,7})$,
    $(A',Z',\zeta') = (A_{k,8},Z_{k,8},\zeta_{k,8})$.
    \item The ninth intermediate state is obtained from a squeeze operator, and thus from Eqs. (\ref{eq:soi}-\ref{eq:sof}), with $W = -X_0$.
    The states before and after such application are specified by
    $(A,Z,\zeta) = (A_{k,8},Z_{k,8},\zeta_{k,8})$,
    $(A',Z',\zeta') = (A_{k,9},Z_{k,9},\zeta_{k,9})$.
\end{enumerate}
The final state is finally obtained by modifying the phase factor, according to the equations $\alpha_{k,f}=\alpha_{k,9}$, $z_{k,f}=z_{k,9}$, and $\zeta_{k,f}=\zeta_{k,9}-\sum_{j=1}^N\omega_{\lambda_k,j} t_k/2$.

\subsection{Vibrational response function}

\begin{figure}[h]
\centering
\includegraphics[width=0.22\textwidth]{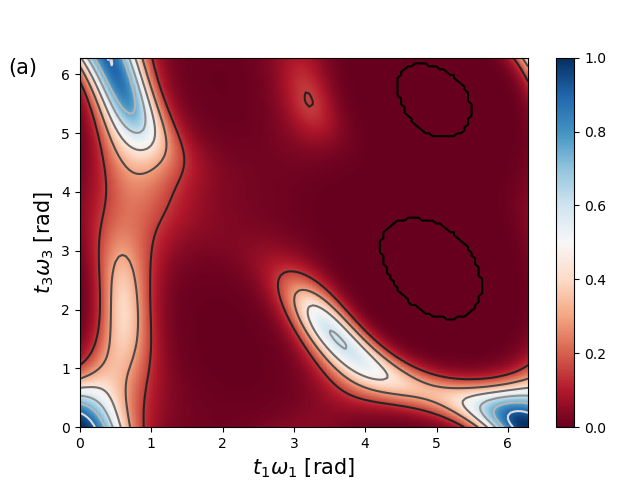}
\includegraphics[width=0.22\textwidth]{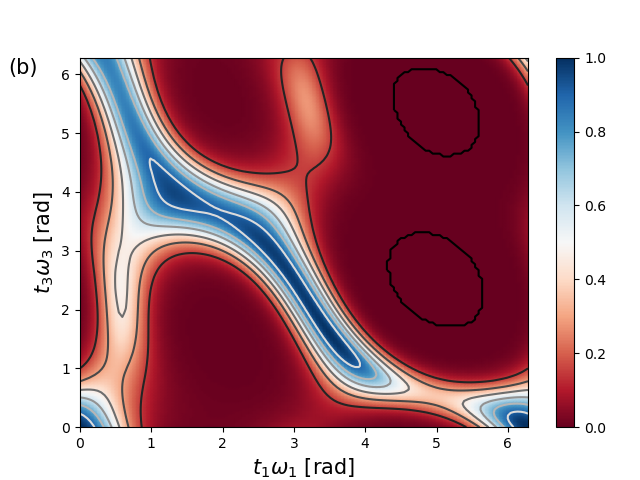}
\includegraphics[width=0.22\textwidth]{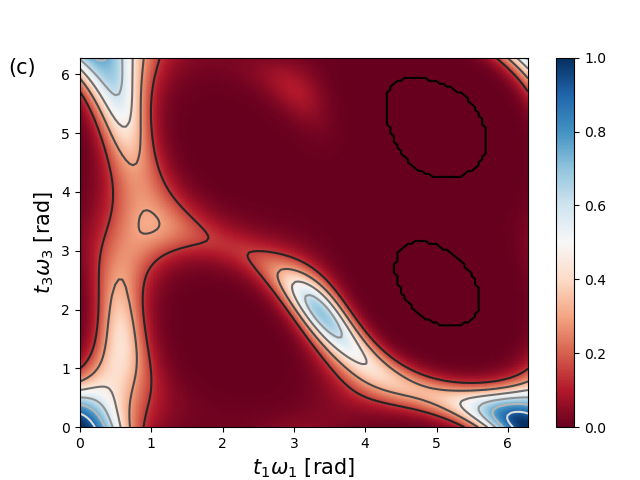}
\includegraphics[width=0.22\textwidth]{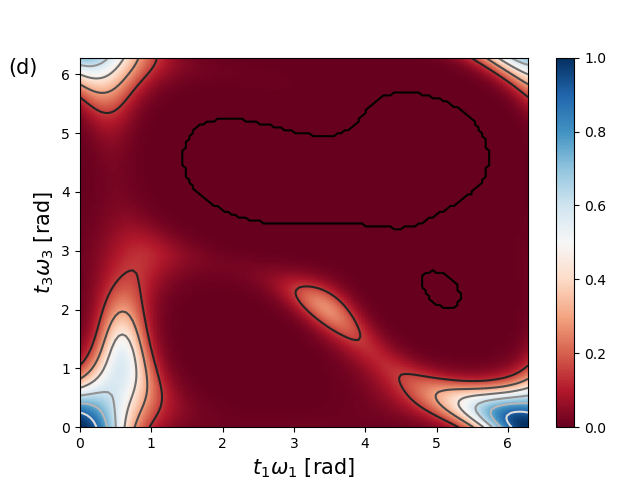}
\includegraphics[width=0.22\textwidth]{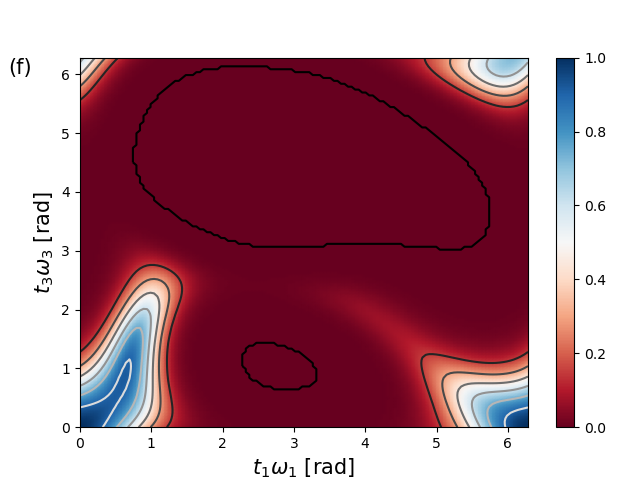}
\includegraphics[width=0.24\textwidth]{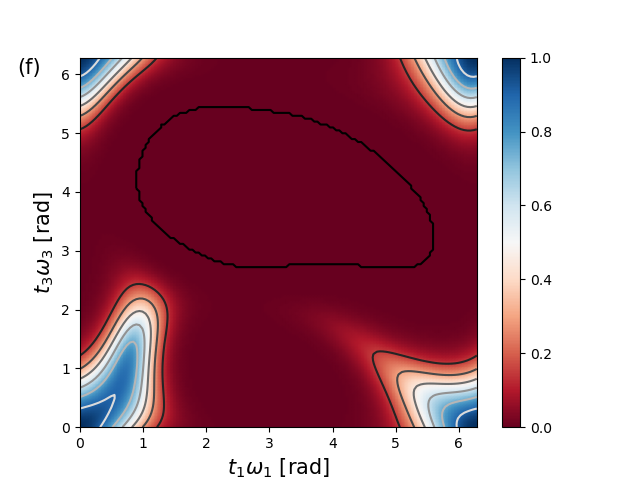}
\caption{Absolute value of the third-order ($M=3$) vibrational response function $R^{(v)}$, as a function of the waiting times $t_1$ and $t_3$, at $t_2\bar\omega_0=1.5\,\pi$. The plots are obtained for $\delta_{1,1}=0.5$ and $\omega_{1,1}/\omega_{0,1}=0.67$ (first mode), $\delta_{1,2}=1.5$ and $\omega_{1,2}/\omega_{0,2}=2$ (second mode). Different panels correspond to different values of the Duschinsky rotation angle $\varphi_1/\pi$, and specifically: (a) $0$, (b) $0.1$, (c) $0.2$, (d) $0.3$, (e) $0.4$, (f) $0.5$.} 
\label{fig:4}
\end{figure}
\begin{figure}[h]
\centering
\includegraphics[width=0.22\textwidth]{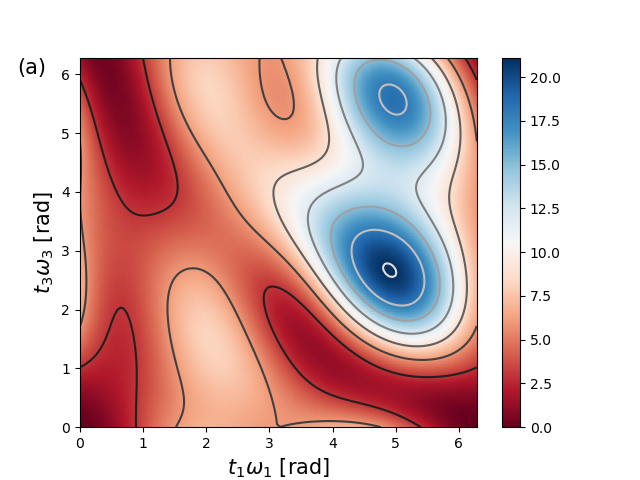}
\includegraphics[width=0.22\textwidth]{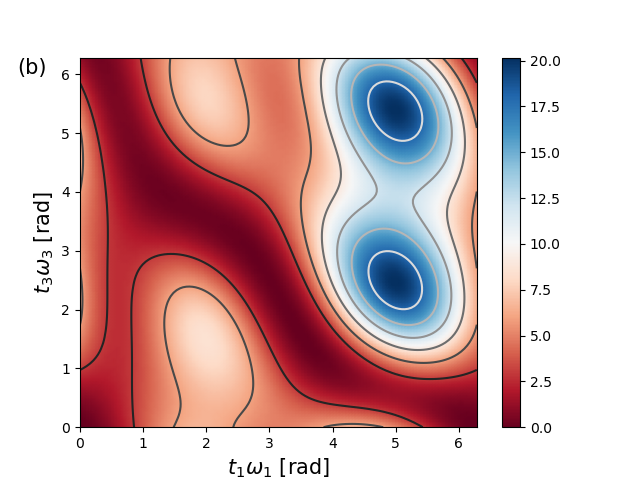}
\includegraphics[width=0.22\textwidth]{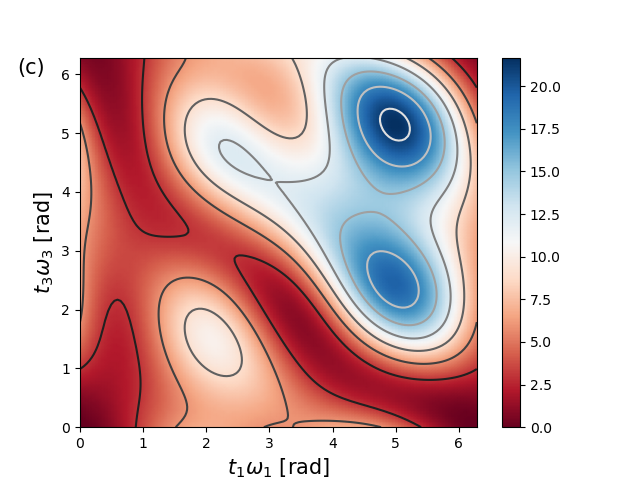}
\includegraphics[width=0.22\textwidth]{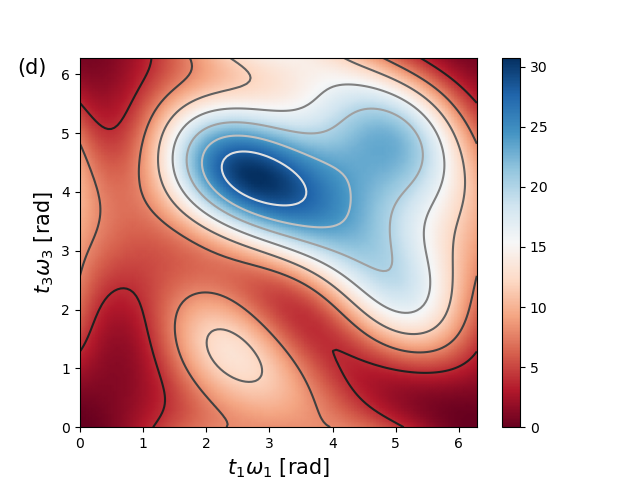}
\includegraphics[width=0.22\textwidth]{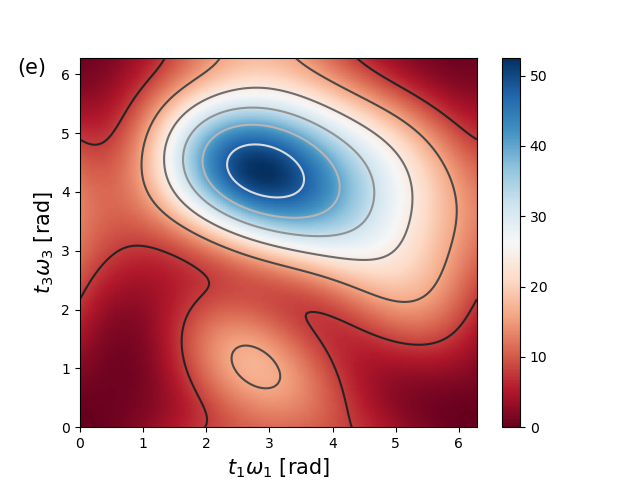}
\includegraphics[width=0.24\textwidth]{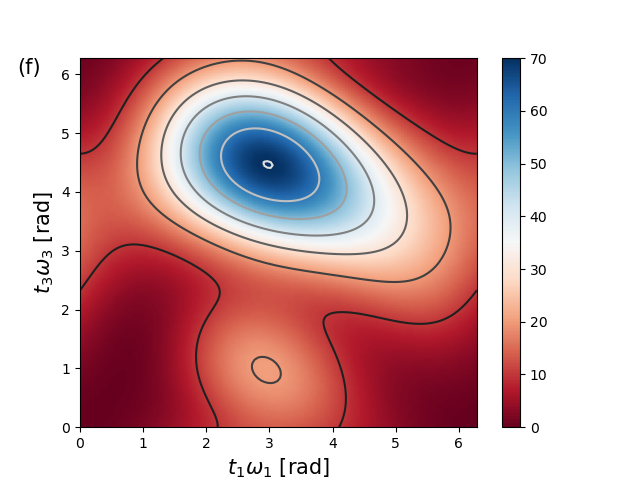}
\caption{Dependence of $|A|^2=|\alpha_1|^2+|\alpha_2|^2$ on the waiting times $t_1$ and $t_3$, for $t_2\bar\omega_0=1.5\,\pi$. The plots are obtained for $\delta_{1,1}=0.5$ and $\omega_{1,1}/\omega_{0,1}=0.67$ (first mode), $\delta_{1,2}=1.5$ and $\omega_{1,2}/\omega_{0,2}=2$ (second mode). Different panels correspond to different values of the Duschinsky rotation angle, and specifcailly: $\varphi_1/\pi$: (a) $0$, (b) $0.1$, (c) $0.2$, (d) $0.3$, (e) $0.4$, (f) $0.5$.} 
\label{fig:5}
\end{figure}
The vibrational component of the $M$-th order response function is given by the overlap between the final vibrational state and the initial ground state, i.e. to that between a vacuum and a squeezed coherent state \cite{Xin89a}:
\begin{gather}
    R^{(v,M)}_\Lambda (t_1,\dots,t_M) = e^{i\zeta} \langle 0 | A, Z \rangle =\nonumber\\ \label{eq:fff}
e^{i\zeta} d_{S_Z}^{1/2} \,\exp\left[-\frac{1}{2} (A^\dagger A -A^\dagger T_Z A^*)\right]
\end{gather}
where $d_{S_Z}$ is the determinant of $S_Z$, while the squeezed coherent state is given by $A=A_{M,f}$, $Z=Z_{M,f}$, and $\zeta = \zeta_{M,f} + \sum_{j=1}^N \omega_{0,j} \sum_{k=1}^M t_k/2 $. As to the matrices, given the polar decomposition $Z=|Z|e^{i\Theta_Z}$, they are defined by the relations $T_Z=\tanh |Z| e^{i\Theta_Z}$ and $S_Z={\rm sech} |Z|$. 

If the initial state of the vibrational modes is a generic squeezed coherent state, $|A_0,Z_0\rangle$, the derivation of the response function must include an additional rotation, squeeze and displacement operator. Equation (\ref{xyz}) applies, as in the single-mode case, but the fictitious final state $| \psi_{ket,\Lambda}' \rangle$ is now given by 
\begin{gather}\label{eq:51u}
    |\psi_{ket,\Lambda}'\rangle = S_N(-Z_0)\, D_N(-A_0)\, R_N(\Phi_0) \nonumber\\ \left(e^{-i H_{v,\lambda_M}t_M/\hbar} \dots e^{-i H_{v,\lambda_1}t_1/\hbar}\right) | A_0 , Z_0 \rangle ,
\end{gather}
where
$\Phi_{0} = \sum_{j=1}^M t_j \, {\rm diag} (\omega_{0,1},\dots,\omega_{0,N})$,
and the product of time evolution operators in parentheses is reduced to the product of squeeze, displacement and rotation operators as shown above.

\subsection{Application to some representative case}

As an illustrative example, we apply the above results to the case of linear and nonlinear response functions for a system with two electronic levels ($N_e=2$) and two vibrational modes ($N=2$). 
Given the presence of only two vibrational modes, the Duschinsky matrix can be expressed as a function of a single rotation angle \cite{Doktorov75a}:
\begin{gather}
    U_\lambda =
    \left(
    \begin{array}{cc}
         \cos\varphi_\lambda & \sin\varphi_\lambda \\
        -\sin\varphi_\lambda & \cos\varphi_\lambda
    \end{array}
    \right).
\end{gather}
The corresponding rotation operator $R_2(\Phi_\lambda)$ is defined by the imaginary and Hermitian matrix 
\begin{gather}
    \Phi_\lambda =
    \left(
    \begin{array}{cc}
         0 & -i\varphi_\lambda \\
        i\varphi_\lambda & 0
    \end{array}
    \right).
\end{gather}
Given the presence of only two electronic states, there is only one relevant rotation operator $R_2(\Phi_1)$, defined by the above expression, with a rotation angle $\varphi_1$.

\subsubsection{First order}

In the linear case ($M=1$), the electronic component of the response function is given by Eq. (\ref{eq:erf1}). The vibrational response function is obtained from the  equations presented in this Section, for $M=1$ and $\lambda_1=1$. In the absence of a rotation ($\varphi_1=0$), the final two-mode state is simply given by the product of two single-mode squeezed coherent states, and $R^{(v,1)}_\Lambda(t_1)$ coincides with the product of the corresponding single-mode response functions. The trajectories described by $\alpha_1$ and $\alpha_2$ are periodic and independent from one another, with periods $2\pi/\omega_{1,1}$ and $2\pi/\omega_{1,2}$, respectively [Fig. \ref{fig:3}(a,b), brown curves]. For $\varphi_1\neq 0$, the Duschinsky rotation couples the two frequencies and complicates the trajectories, whose period now coincides with that of the overall rotation operator $R_2(\Phi_1')$. This is given by $T=2k_1\pi/\omega_{1,1}=2k_2\pi/\omega_{1,2}$, being $k_1$ and $k_2$ the smallest integers such that fulfil the second equality (in the reported example, $k_1=11$ and $k_2=9$, corresponding to $T\bar\omega=20\pi$). If the frequencies $\omega_{1,1}$ and $\omega_{1,2}$ are not commensurate, then the trajectories described by the $\alpha_j$ in the respective planes are in general open.

The vibrational response function presents features that are qualitatively similar to those observed in the single-mode case, but with a longer period [Fig. \ref{fig:3}(c)] or with no periodic behavior, depending on the ratio between the two excited-state frequencies $\omega_{1,j}$ ($j=1,2$). Comparing the time dependence of $\alpha_1$ and $\alpha_2$ with that of the response function's modulus, it emerges that its main maxima correspond to the times where both vibrational wave packets are close to the origin ({\it i.e.}, both $\alpha_1$ and $\alpha_2$ are close to zero). The lower maxima, instead, correspond to times where only one of the two wave packets is close to the origin.

\subsubsection{Third-order}

For the third-order term ($M=3$), we specifically consider the non-rephasing part of the ground-state bleaching contribution [$\lambda_1=\lambda_3=1$ and $\lambda_2=0$, Fig. \ref{fig:0}(b)]. The electronic part of the response function is given in Eq. (\ref{eq:erf3}).
The vibrational response function is displayed in Fig. \ref{fig:4} for different values of the mixing angle $\varphi_1$. Its periodicity in the waiting times $t_1$ and $t_3$ follows the same properties outlined above for the first-order response function. The mixing angle changes the functional dependence of the response function on the waiting times $t_1$ and $t_3$. In analogy to what happens in the single-mode case, such dependence is clearly correlated to that of $|A|^2$ (Fig. \ref{fig:5}), being $|A,Z\rangle$ the final state of the two vibrational modes. This results from the fact that the overlap between $|A,Z\rangle$ and the initial vibrational (ground) state, exponentially depends on the the square modulus of $A$ [see Eq. (\ref{eq:fff})]. Even though the combination of rotation, squeeze and displacement operators gives rise to nontrivial two-mode states, the connection between the vibrational response function and the position of the vibrational wave packets thus remains clear.

\section{Conclusions}

In conclusion, we develop an approach for the calculation of the vibrational response functions to be applied in the simulation of multidimensional electronic spectroscopy. The approach applies to systems whose $N$ (relevant) vibrational modes can be described as independent harmonic oscillators, whose frequencies and origin depend on the electronic state. Besides, the field is assumed to induce Franck-Condon transitions between electronic states, resulting in a mixing of the normal vibrational modes (Duschinsky rotations). The approach allows the explicit derivation of the squeezed coherent multi-mode vibrational state generated by the sequence of field-induced excitations and free-evolution periods, whose overlap with the initial state gives the vibrational response function. From both the formal and the numerical points of view, the calculation of the response function at given waiting times is obtained by applying simple {\color{black} iterative} relations (nine for each time interval) to the three quantities that define the vibrational state: a complex $N$-dimensional vector $A$, a complex $N \times N$ matrix $Z$, and a real number $\zeta$. This replaces the time integration of the Schr\"odinger equation or the Hamiltonian diagonalization, combined with the sum over infinite vibronic pathways, that are required in alternative approaches. From a physical point of view, the present allows the derivation of the vibrational state generated within each electronic pathway, thus relating a clear physical picture to the behavior of the vibrational response function.

\acknowledgements

This work has been supported by the European Union’s Horizon 2020 research and innovation programme under  the  Marie  Sklodowska-Curie  Grant  Agreement No. 812992.

\appendix 

{\color{black} 
\section{Properties of the response functions \label{app:z}}
The contributions to the response function related to each electronic pathway can be written in a factorized form. Also, vibrational response functions that involve the same sequence of electronic states but correspond to different kinds of pathways can be derived from one another by suitable redefinitions of the waiting times. These properties are demonstrated in the following two paragraphs.
\paragraph{Factorization of the response function.}
The third-order response function corresponds to the four-point correlation function reported in Eq. \ref{eq:sdd}. Replacing each dipole operator with its expression in terms of the time-evolution operators, one obtains terms such as:
\begin{gather}
    \langle \mu (t_{123})\,\mu (t_{12})\,\mu (t_1)\,\mu\,\rho_0\rangle \nonumber\\ 
    =\langle 0 | \, U^\dagger(t_{123})\, \mu\, U(t_3)\, \mu\, U(t_2)\, \mu\, U(t_1)\, \mu\, |0\rangle \nonumber\\
    =(i/\hbar)^{-3}\sum_{\Lambda} R^{(e,3)}_{\Lambda}\,R^{(v,3)}_{\Lambda} ,
\end{gather}
which specifically corresponds to the term in Eq. (\ref{eq:sdd}) where all the dipole operators act on the density operator from the left. The vector ${\Lambda} =(\lambda_1,\lambda_2,\lambda_3)$ specifies the electronic states obtained after the application to the ket of the first, second, and third dipole operator, respectively. 
The possibility of writing each term in the above sum in a factorized form results from the expression of the time evolution operator related to the Hamiltonian in Eq. (\ref{eq:ham1}):
\begin{gather}
    U = \sum_\lambda U_{e,\lambda} \otimes U_{v,\lambda} = \sum_\lambda |\lambda\rangle\langle\lambda| e^{-i\epsilon_\lambda t/\hbar} \otimes e^{-i H_{v,\lambda} t /\hbar} .
\end{gather}
In particular, the electronic component of the response function reads:
\begin{gather}
R^{(e,3)}_{\Lambda} = C_{\Lambda} e^{-i(\epsilon_{\lambda_3} t_3+\epsilon_{\lambda_2} t_2+\epsilon_{\lambda_1} t_1)/\hbar} ,
\end{gather}
where $C_{\Lambda}=(i/\hbar)^3\,\mu_{\lambda_1 0}\,\mu_{\lambda_2\lambda_1}\,\mu_{\lambda_3\lambda_2}\,\mu_{0\lambda_3}$.
The vibrational component of the response function is given by:
\begin{gather}
R^{(v,3)}_{\Lambda} = \langle 0_0 |\, U_{v,0}^\dagger (t_{123})\, U_{v,\lambda_3} (t_{3})\, U_{v,\lambda_2} (t_{2})\, U_{v,\lambda_1} (t_{1})\, | 0_0 \rangle \nonumber\\ \label{eq:ffdfd}
= e^{it_{123}\omega_{0}/2} \langle 0_0 |\, U_{v,\lambda_3} (t_{3})\, U_{v,\lambda_2} (t_{2})\, U_{v,\lambda_1} (t_{1})\, | 0_0 \rangle ,
\end{gather}
where $|n_\lambda\rangle = |0_0\rangle$ is the ground state of the undisplaced harmonic oscillator. 

\paragraph{Terms related to different processes.}
The contribution to the response function explicitly derived in the above paragraph corresponds to the case where all the dipole operators act on the left side of the density operator $\rho_0$. In the case $\lambda_2=0$, this would be the non-rephasing component of a ground state bleaching contribution. Let's now consider, for example, the case where the dipole operators $\mu$ and $\mu(t_{12})$ in Eq. (\ref{eq:sdd}) act on the right of $\rho_0$:
\begin{gather}
    \langle \mu (t_{123})\,\mu (t_1)\,\rho_0\,\mu\,\mu (t_{12})\,\rangle =
    \langle \mu\,\mu (t_{12})\,\mu (t_{123})\,\mu (t_1)\,\rho_0\rangle 
    \nonumber\\ 
    =\langle 0 | \, \mu\, U^\dagger(t_{12})\, \mu\, U^\dagger (t_3)\, \mu\, U(t_{23})\, \mu\, U(t_1)\,  |0\rangle \nonumber\\
    =(i/\hbar)^{-3}\sum_\Lambda R^{(e,3)}_{\Lambda,\Gamma}\,R^{(v,3)}_{\Lambda,\Gamma},
\end{gather}
where in the first equation we have exploited the cyclic property of the trace. The vector $\Gamma = (L,R,L)$ specifies the side of the double-sided Feynman diagram where the first three interactions with the field take place (it is intended that, when the vector is omitted, all interactions take place on the left side). For $\lambda_2=0$, this would be the rephasing component of a stimulated emission contribution. Proceeding as in the previous paragraph, one obtains the following expression for the electronic component of the response function:
\begin{gather}
R^{(e,3)}_{\Lambda,\Gamma} = C_{\Lambda,\Gamma} e^{i(\epsilon_{\lambda_3} t_{12}+\epsilon_{\lambda_2} t_3-\epsilon_{\lambda_1} t_{23})/\hbar} ,
\end{gather}
where $C_{\Lambda,\Gamma}=(i/\hbar)^3\,\mu_{j0}\,\mu_{kj}\,\mu_{lk}\,\mu_{0l}$.
The vibrational component of the response function is given by:
\begin{gather}
R^{(v,3)}_{\Lambda,\Gamma} = \langle 0_0 |\, U_{v,\lambda_3}^\dagger (t_{12})\, U^\dagger_{v,\lambda_2} (t_{3})\, U_{v,\lambda_1} (t_{23})\, U_{v,0} (t_{1})\, | 0_0 \rangle \nonumber\\ \label{eq:dedw}
= e^{-it_{1}\omega_{0}/2} \langle 0_0 |\, U_{v,\lambda_3}^\dagger (t_{12})\, U^\dagger_{v,\lambda_2} (t_{3})\, U_{v,\lambda_1} (t_{23})\, | 0_0 \rangle .
\end{gather}
From a comparison between Eq. (\ref{eq:ffdfd}) and Eq. (\ref{eq:dedw}) it follows that
\begin{gather}
    R_{\Lambda,\Gamma}^{(v,3)}(t_1,t_2,t_3) = R_{\Lambda}^{(v,3)}(t_{23},-t_3,-t_{12}),
\end{gather}
with $\Gamma=(L,R,L)$.

One can show that analogous relations can be established between contributions characterized by the same sequences of electronic states in the double-sided Feynman diagram, moving clockwise from the first state on the left (ket) to the last one on the right (bra). In all cases, for a generic $\Lambda$, you end up with a response function of the form
\begin{gather}
R^{(v,3)}_{\Lambda,\Gamma}  
= e^{it_{abc}\omega_{0}/2} \langle 0_0 |\, U_{v,\lambda_3} (t_c)\, U_{v,\lambda_2} (t_b)\, U_{v,\lambda_1} (t_a)\, | 0_0 \rangle \nonumber,
\end{gather}
where $t_{abc}=t_a+t_b+t_c$ and $t_a$, $t_b$, $t_c$ are given by different combinations of $t_1$, $t_2$, $t_3$ \cite{Troiani23a}.
}

\section{Properties of the single-mode rotation, displacement, and squeeze operators \label{app:a}}

The time evolution generated by the vibrational Hamiltonians $H_{v,\lambda}$ [Eq. (\ref{eq:hamvib})] can be derived from the properties of the rotation, displacement, and squeeze operators \cite{Scully97a}, which are recalled hereafter.

\paragraph{Rotation operator.}
The rotation operator can be expressed as an exponential function of the number operator $n=a^\dagger a$:
\begin{gather}
    R(\phi) = \exp\left(i\phi\, a^\dagger a\right) ,
\end{gather}
where $\phi$ is real and defines the rotation angle.
It's a unitary operator, whose inverse is given by:
\begin{gather}
    R^{-1} (\phi) = R^\dagger (\phi) = R(-\phi) .
\end{gather}
Finally, the product of two rotations is still a rotation operator:
\begin{gather}\label{C1}
    R(\phi_1)\,R(\phi_2)=R(\phi_3),
\end{gather}
where $\phi_3=\phi_1+\phi_2$.

\paragraph{Displacement operator.}
The displacement operator is defined can be written as an exponential function of the creation and annihilation operators:
\begin{gather}\label{A03q}
    D(\alpha) = \exp\left(\alpha a^\dagger - \alpha^* a\right) ,
\end{gather}
where $\alpha$ is a complex number. 
The displacement operator is unitary, and its inverse is given by:
\begin{gather}\label{A04}
    D^{-1} (\alpha) = D^\dagger (\alpha) = D(-\alpha) .
\end{gather}
The product of two displacement operators is still a displacement operator, times a phase factor:
\begin{gather}\label{C2}
    D(\alpha_1)\,D(\alpha_2)=e^{(\alpha_1\alpha_2^*-\alpha_1^*\alpha_2)/2}D(\alpha_1+\alpha_2).
\end{gather}

\paragraph{Squeeze operator.}
The squeeze operator is given by an exponential function of the creation and annihilation operators squared:
\begin{gather}\label{A06q}
    S(z) = \exp\left\{\frac{1}{2}\left[z^*a^2-z\left(a^\dagger\right)^2\right]\right\},
\end{gather}
where $z$ is a complex number.
This operator is also unitary, and its inverse is given by:
\begin{gather}\label{A07}
    S^{-1} (z) = S^\dagger (z) = S(-z) .
\end{gather}
Finally, the product of two squeeze operators corresponds to that of a squeeze and a rotation operator:
\begin{gather}\label{C3}
    S(z_1)\,S(z_2) = e^{i\phi/2} S(z_3)\,R(\phi)
\end{gather}
where $z_j \equiv |z_j| e^{i\theta_j}$ and ($j=1,2,3$).
The final squeezing parameters and rotation angle are defined by
\begin{gather}
    t_3 = \frac{t_1+t_2}{1+t_2 t_1^*} ,\ 
    e^{i\phi} = \frac{1+t_1 t_2^*}{\left|1+t_1 t_2^*\right|},\label{Z2}
\end{gather}
with $t_j\equiv e^{i\theta_j}\tanh |z_j|$.

\paragraph{Commutation relations.}

Rotation, displacement, and squeeze operators don't commute with each other.
This can be deduced from the effect of these transformations on the annihilation operator, which is given by the following equations:  
\begin{gather}
R(-\phi)\,a\,R(\phi)=a\,e^{i\phi}\\
S(-z)\, a \, S(z) = a\,\cosh |z| - a^\dagger\,e^{i\theta}\,\sinh |z| \\
D(-\alpha)\, a \, D(\alpha) = a + \alpha
\end{gather}
Combining the above equations with the definitions of the displacement, rotation, and squeeze operators, one obtains the expressions: 
\begin{gather}
    D(\alpha)\,R(\phi) = R(\phi)\, D(\alpha e^{-i\phi}) \label{B1}\\
    S(z)\,R(\phi) = R(\phi)\, S(z e^{-i2\phi})  \label{B2} \\
    S(z)\,D(\alpha)=D(\alpha\cosh |z|-\alpha^*e^{i\theta}\sinh |z|)\,S(z) \label{B3}.
\end{gather}

\paragraph{Application to the squeezed coherent states.}

From the above combination and commutation relations it follows that the application of a rotation, displacement, or squeeze operator to a squeezed coherent state produces another squeezed coherent state.

Let's start by considering the application of a rotation operator. This gives:
\begin{gather}
    R(\phi)\,|\alpha,z\rangle = D(e^{i\phi}\alpha)\,R(\phi)\,S(z)\,|0\rangle = \nonumber\\ 
    D(e^{i\phi}\alpha)\,S(e^{2i\phi}z)\,|0\rangle = |e^{i\phi}\alpha,e^{2i\phi}z\rangle .
\end{gather}

As to the displacement operator, by applying the combination of the displacement operators, one obtains:
\begin{gather}
D(\beta)\,|\alpha,z\rangle = e^{(\alpha^*\beta-\alpha\beta^*)/2}  D(\alpha+\beta)\,S(z)\,|0\rangle \nonumber\\ 
    = e^{(\alpha^*\beta-\alpha\beta^*)/2} |\alpha+\beta,z\rangle .
\end{gather}

Finally, the application of a squeeze operator to a squeezed coherent state is given by the equation:
\begin{gather}
S(w)\,|\alpha,z\rangle = D(\beta)\,S(w)\,S(z)\,|0\rangle = e^{i\phi/2}\, |\beta,u\rangle ,
\end{gather}
where 
\begin{gather}
 \beta=\alpha\cosh |w|-\alpha^*e^{i\theta_z}\sinh |w|   
\end{gather}
and
\begin{gather}
t_u = \frac{t_w+t_z}{1+t_w^*\,t_z},\
    e^{i\phi} = \frac{1+t_w\,t_z^*}{|1+t_w\,t_z^*|}.
\end{gather}

\section{Properties of the multi-mode rotation, displacement, and squeeze operators \label{app:b}}

The time evolution generated by the vibrational Hamiltonians $H_{v,\lambda}$ [Eq. (\ref{eq:mmvh})] can be derived from the properties of the multimode rotation, displacement, and squeeze operators \cite{Xin90a}, which are recalled hereafter.

\paragraph{Rotation operator.}
The rotation operator can be expressed as an exponential function of the vectorial creation and annihilation operators, $\tilde a=(a_1,\dots,a_N)$ and $\tilde a^\dagger=(a_1^\dagger,\dots,a_N^\dagger)$ {\color{black} (the tilde denotes the transposition of the matricial operator)}:
\begin{gather}
    R_N(\Phi) = \exp\left(i \tilde a^\dagger\,\Phi\, a\right) ,
\end{gather}
where $\Phi$ is a Hermitian matrix.
It's a unitary operator, whose inverse is given by:
\begin{gather}
    R_N^{-1} (\Phi) = R_N^\dagger (\Phi) = R_N(-\Phi) .
\end{gather}
Finally, the product of two rotations is still a rotation operator:
\begin{gather}\label{C1p}
    R_N(\Phi_1)\,R_N(\Phi_2)=R_N(\Phi_3),
\end{gather}
where
$e^{i\Phi_3}=e^{i\Phi_1} e^{i\Phi_2}$.

\paragraph{Displacement operator.}
The displacement operator can be written as an exponential function of the creation and annihilation operators:
\begin{gather}\label{A03r}
    D_N(A) = \exp\left(\tilde A a^\dagger - A^\dagger a\right) ,
\end{gather}
where $\tilde A=(\alpha_1,\cdots,\alpha_N)$ and $A^\dagger=(\alpha_1^*,\cdots,\alpha_N^*)$ are a complex vectors {\color{black} (the tilde and the dagger applied to a numerical matrix denote its simple and conjugate transpositions, respectively)}. 
The displacement operator is unitary, and its inverse is given by:
\begin{gather}\label{A04p}
    D_N^{-1} (A) = D_N^\dagger (A) = D_N(-A) .
\end{gather}
The product of two displacement operators is still a displacement operator, times a phase factor:
\begin{gather}\label{C2p}
    D_N(A_1)\,D_N(A_2)=e^{(A_2^\dagger A_1 -A_1^\dagger A_2)/2}\, D_N(A_1+A_2),
\end{gather}
where $A_i^\dagger A_j$ are matrix products, whose result corresponds to scalars.

\paragraph{Squeeze operator.}
The squeeze operator is given by an exponential function of the creation and annihilation operators squared:
\begin{gather}\label{A06r}
    S_N(Z) = \exp\left[\frac{1}{2}\left(\tilde a Z^\dagger a -\tilde a^\dagger Z a^\dagger\right)\right],
\end{gather}
where $Z$ is a symmetric matrix.
This operator is also unitary, and its inverse reads:
\begin{gather}\label{A07p}
    S_N^{-1} (Z) = S_N^\dagger (Z) = S_N(-Z) .
\end{gather}
Finally, the product of two squeeze operators corresponds to that of a squeeze and a rotation operator:
\begin{gather}\label{C3P}
    S_N(Z_1)\,S_N(Z_2) = e^{\frac{i}{2}{\rm Tr}(\Phi)} S_N(Z_3)\,R_N(\Phi)
\end{gather}
where $Z_j \equiv |Z_j| e^{i\Theta_j}$  ($j=1,2,3$), $|Z_j|=\sqrt{Z_j^\dagger Z_j}$, and 
\begin{gather}\label{C4P}
T_3=S_1^{-1} (T_1+T_2) (I+T_1^\dagger T_2)^{-1} \tilde S_1\\
e^{i\Phi} = S_3^{-1} S_1 (I+T_2 T_1^\dagger)^{-1} S_2
\end{gather}
being $ T_j \equiv \tanh |Z_j|\,e^{i\Theta}$ and $S_j={\rm sech}\, |Z_j|$.

\paragraph{Commutation relations.}

Rotation, displacement, and squeeze operators don't commute with each other.
This can be deduced from the effect of these transformations on the vectorial annihilation operator, which is given by the following equations
\begin{gather}
R_N(-\Phi)\,a\,R_N(\Phi)=e^{i\Phi}\,a\\
S_N(-Z)\, a \, S_N(Z) = \cosh |Z|\, a - \sinh |Z|\, e^{i\Theta}\, a^\dagger \\
D_N(-A)\, a \, D_N(A) = a + A .
\end{gather}
Combining the above equations with the definitions of the displacement, rotation, and squeeze operators, one obtains the expressions: 
\begin{gather}
    D_N(A)\,R_N(\Phi) \!=\! R_N(\Phi)\, D(e^{-i\Phi} A ) \label{B1p}\\
    S_N(Z)\,R_N(\Phi) \!=\! R_N(\Phi)\, S_N(e^{-i\Phi} Z e^{-i\tilde\Phi})\label{B2p}\\
    S_N(Z)\,D_N(A) \!=\! D_N(\cosh |Z|\,A\!-\!\sinh |Z|\,e^{i\Theta}\,A^*)\,S_N(Z) \label{B3p}.
\end{gather}

\paragraph{Application to the squeezed coherent states.}

From the above combination and commutation relations it follows that the application of a rotation, displacement, or squeeze operator to a squeezed coherent state produces another squeezed coherent state.

Let's start by considering the application of a rotation operator. This gives:
\begin{gather}
    R_N(\Phi)\,|A,Z\rangle = D_N(e^{i\Phi}A)\,R_N(\Phi)\,S_N(Z)\,|0\rangle = \nonumber\\ 
    D_N(e^{i\Phi}A)\,S_N(e^{i\Phi} Z e^{i\tilde\Phi})\,|0\rangle = |e^{i\Phi}A,e^{i\Phi} Z e^{i\tilde\Phi}\rangle .
\end{gather}

As to the displacement operator, by applying the combination of the displacement operators, one obtains:
\begin{gather}
D_N(B)\,|A,Z\rangle = e^{(A^\dagger B-B^\dagger A)/2}  D_N(A+B)\,S_N(Z)\,|0\rangle \nonumber\\ 
    = e^{(A^\dagger B-B^\dagger A)/2} |A+B,Z\rangle .
\end{gather}

Finally, the application of a squeeze operator to a squeezed coherent state is given by the equation:
\begin{gather}
S_N(W)\,|A,Z\rangle = D_N(B)\,S_N(W)\,S_N(Z)\,|0\rangle \nonumber\\ = e^{\frac{i}{2}{\rm Tr}(\Phi)}\, |B,U\rangle .
\end{gather}
where 
\begin{gather}
 B=\cosh |W|\,A-\sinh |W|\,e^{i\theta_W}\,A^*   
\end{gather}
and 
\begin{gather}\label{C4Pp}
T_U=S_W^{-1} (T_W+T_Z) (I+T_W^\dagger T_Z)^{-1} \tilde S_W\\
e^{i\Phi} = S_U^{-1} S_W (I+T_Z T_W^\dagger)^{-1} S_Z .
\end{gather}

\bibliography{paper}

\end{document}